\documentclass[12pt,preprint]{aastex}

\def\msun{{\rm ~M}_{\odot}}
\def\rsun{{\rm ~R}_{\odot}}

\usepackage{amsmath}

\begin{document}

\title{The Lack of Gamma-Ray Bursts from Population III Binaries}

 \author{Krzysztof Belczynski\altaffilmark{1,2}, Tomasz
Bulik\altaffilmark{3,4}
         Alexander Heger\altaffilmark{5,6} and Chris
Fryer\altaffilmark{7,8}}

 \affil{
     $^{1}$ New Mexico State University, Dept. of Astronomy,
        1320 Frenger Mall, Las Cruces, NM 88003\\
     $^{2}$ Tombaugh Fellow\\
     $^{3}$ Astronomical Observatory, Warsaw University,
            Al.\ Ujazdowskie 4, 00-478, Warsaw, Poland \\
     $^{4}$ Nicolaus Copernicus Astronomical Center,
            Bartycka 18, 00-716 Warszawa, Poland\\
     $^{5}$ Theoretical Astrophysics, Los Alamos National Laboratory,
            Los Alamos, NM 87545\\
     $^{6}$ Department of Astronomy and Astrophysics, University of
     California, Santa Cruz, CA 95064 \\
     $^{7}$ Computer and Computational Science, Los Alamos National Laboratory,
            Los Alamos, NM 87545\\
     $^{8}$ Physics Department, University of Arizona, Tucson, AZ, 85721 \\
     kbelczyn@nmsu.edu,tb@astrouw.edu.pl,aheger@lanl.gov,fryer@lanl.gov}

 \begin{abstract}
We study the evolution of first star (Population III) binaries. Under 
specific conditions, these stars may produce high redshift gamma-ray
bursts (GRBs). We demonstrate that the occurrence rate of GRBs does
not depend sensitively on evolutionary parameters in the population
synthesis models. We show that the first binaries may form a very small
group ($\lesssim 1\%$) of fast rotating stars through binary tidal
interactions that make GRBs.  This finding is contrary to the Bromm
\& Loeb assumption that {\em all} stars in close Population III binaries 
will be spun up by tides and produce a GRB.  We find that there is simply
not enough fast rotating stars in Population III binaries to expect
detection with {\em SWIFT}. Predicted detection rates, even with very
optimistic assumptions on binary fraction, evolutionary parameters and
GRB detection, are very small: $~\sim 0.1-0.01 {\rm yr}^{-1}$.
 \end{abstract}


\section{Introduction}

Population III stars are formed out of a pristine, metal free gas made
in the big bang.  These are the first stars in the universe and most
of these stars have likely formed at a redshift of $z \gtrsim 10$.  To
collapse and form stars in this metal-free environment, with its
cooling dominated by H$_2$, clouds must be larger than in metal-rich
environs. Current simulations indicate that an
initial mass function of Population III stars is expected to be
top heavy with a large fraction of the mass above $100\,M_\odot$
(Larson 1998; Nakamura \& Umemura 2001; Abel, Bryan, \& Norman 2002;
Bromm, Coppi, \& Larson 2002; Chabrier 2003), though somewhat less
massive, but still massive, stars could form in relic HII regions
(O'Shea et al.\ 2005).  Metal free stars also have much weaker (if
winds) and hence, much less mass loss than current stellar populations
(Kudritzki 2002) and do not show strong pulsational instabilities that
could lead to strong mass loss for initial masses up to
500-1000\,M$_\odot$ (Baraffe, Heger \& Woosley 2001).  That is, not
only are these stars born massive, but they also keep most of their
mass till their late evolution stages, and a large fraction of them
form black holes. There are also calculations that indicate that these 
stars could have mass loss originating from high rotation (e.g., Marigo
2001). 

The evolution of metal free stars has been investigated by several
authors (e.g., Woosley, Heger, \& Weaver 2002; Chieffi \& Limongi
2002; Umeda \& Nomoto 2002).  For Population III stars, the pair
instability is important, and it leads to a complete disruption of
stars with initial masses between 140 and 260 \,M$_\odot$ (Bond et al.
1984;  Heger \& Woosley 2002).  Below this mass range, pair
instabilities lead to pulsations and some mass loss, while above
260\,M$_\odot$, the stars collapse directly to form black holes
(Fryer, Woosley \& Heger 2001).  Population III stars are a primary
source of ionizing photons in the early universe and, if they explode,
can very quickly enrichment the interstellar medium with metals.  The
traces of this ionization have been measured by the Wilkinson
Microwave Anisotropy Probe (Kogut 2003; Cen 2003; Wyithe \& Loeb
2003a, 2003b).

As every known stellar population contains a significant fraction of
binary and multiple stars, it is interesting to ask if it was the case
for the Population III stars as well. The results of numerical
simulation show that metal free clouds with significant angular
momentum form binaries (Saigo, Matsumoto \& Umemura 2004). Thus it is
likely that Population III stars formed in binaries, however, this
issue is still a subject of discussion (see Ripamonti \& Abel
2004). The evolution of Population III binaries has been first
addressed by Belczynski, Bulik \& Rudak (2004), who considered the
possibility of forming double black hole binaries as a result of
Population III binary evolution. The coalescences of such binaries may
be detectable by gravitational wave detectors such as LIGO and VIRGO
(Kulczycki et al. 2006). Formation of double black hole binaries from
single Population III stars through dynamical interactions and
corresponding LIGO signal was also considered (Wyithe \& Loeb 2004).

Gamma-ray bursts may be a different observational manifestation of
Population III stars.  The observed population of long soft GRBs
extends to the redshifts above 6.  GRBs from higher redshifts are
expected to be detectable by current instruments, and so are their
afterglows (Lamb \& Reichart 2000).  The collapsar model (Woosley
1993; MacFadyen \& Woosley 1999) seems to match many of the observed
qualities of long duration GRBs: burst location in star-forming
galaxies and/or star-forming regions in galaxies (Fruchter et
al. 1999; Vreeswijk et al. 2001) and the concurrent production of
supernovae in the GRB explosion (e.g., Galama et al. 1998).  
Moreover, GRBs host galaxies tend to be metal poor (Stanek et al. 2006).
The main requirements in the collapsar scenario are that {\em (i)} the star
loses its extended envelope, so the relativistic jet can punch through the
compact core of a collapsing star on a timescale comparable to a typical
timescale of long-soft GRB (Fryer, Woosley \& Hartmann 1999),
{\em (ii)} the star retains, or attains, a significant amount of
angular momentum, so that a self-supporting torus may form around a
black hole during core collapse. Given the high masses of Population
III stars at collapse, one might expect Population III stars to
produce a lot of GRBs.  However, if Population III stars lose no or
only very little mass in stellar winds\footnote{Although this would
seem to rule out single-star progenitors, stellar evolution theorists
have found that rotation-induced mixing can increase the helium layer
for fast-spinning single stars, allowing single Population III stars
to produce GRBs (Woosley \& Heger 2006; Yoon \& Langer 2005).}, they
retain their hydrogen envelopes and hence do not satisfy requirement
{\em (iii)}; (see also Heger et al.\ 2003). Binary evolution can both
help remove the hydrogen envelope (through stripping in Roche lobe
overflow) and spin-up binary components.

The binary scenario for Population III GRBs has been considered by
Lloyd-Ronning et al. (2002) who studied the dependence of GRB
formation on metallicity and later by Bromm \& Loeb (2006) who
calculated the SWIFT detection rate using a simplified analytical
approach to binary evolution.  Although Lloyd-Ronning et
al. (2002) found only mild or no increase in the ratio of GRBs to SNe
for Population III stars, Bromm \& Loeb (2006) argued that Population III
stars may significantly enhance (in addition to Population I/II
progenitors) the rate of GRBs observed at high redshifts.  However,
simplifications in both Lloyd-Ronning et al. (2002) and Bromm \&
Loeb (2006) evolutionary models (e.g., they did not calculate the
angular momentum of the collapsing star) limited the extent at which
these calculations could really estimate the GRB rate.

In this paper we investigate evolutionary scenarios leading to
Population III collapsars in binary systems focusing on progenitors
spin evolution. We use detailed stellar models for the evolution of
Population III stars and follow some basic
binary processes important for the angular momentum evolution of potential
GRB progenitors.  The paper is organized as follows: in \S\,2
we present the evolutionary model of Population III stars (singles and
binaries), in \S\,3 we describe the results and Population III
GRB rates, and \S\,4 contains our conclusions and a summary.

\section{Model Description}

To  follow the evolution of Population III single stars we use models
from Heger \& Woosley (2006) for 10-100 solar masses, Scannapieco et al.
(2005) for more massive stars, and additionally models for lower mass
stars (which extend to late evolutionary stages, i.e., AGB) and some high
mass stars ($100-500 \msun$) specifically calculated for this study. The
revised model of Belczynski et al. (2004) is used to evolve Population III
binaries. Final black hole masses and formation of collapsar GRBs are
obtained from Fryer (2006) semi-analytic estimates based on numerical
simulations.

\subsection{Single Star Evolution}

The single star evolution was followed using the KEPLER code (Weaver,
Zimmerman, \& Woosley 1978), an implicit 1D hydro code. A set of published
models was used (Heger \& Woosley 2007) and additionally some lower mass
($<10 \msun$) and higher mass ($> 100 \msun$) models were calculated.
We use an initial primordial composition and assume no mass loss.  The
models
used here do not include effects of rotation on the evolution of a star;
the initial angular momentum of Population III binaries is not known.  
The assumptions for semi-convection are such that the AGB-core collapse
transition (off-center neon ignition) occurs at $\gtrsim 10 \msun$,
and the first noticeable pulsational pair instability sets in just below
$100 \msun$.  One of the big uncertainties of these models is the question
of mixing of carbon, made in the convective core during core helium
burning, with the hydrogen envelope (see, e.g., Heger, Woosley, Waters
2000; Marigo et al.\ 2001). The massive star models used here show only
moderate mixing and remain relatively compact. However, study of
Scannapieco et al. (2005), who varied assumptions on semi-convection and
overshoot mixing, showed that a large range of radii at the end of core
helium burning is possible for massive stars ($\sim 150-250 \msun$).
This constitutes a major uncertainty for Population III stellar models.

\begin{figure}
\includegraphics[width=0.9\columnwidth,angle=0]{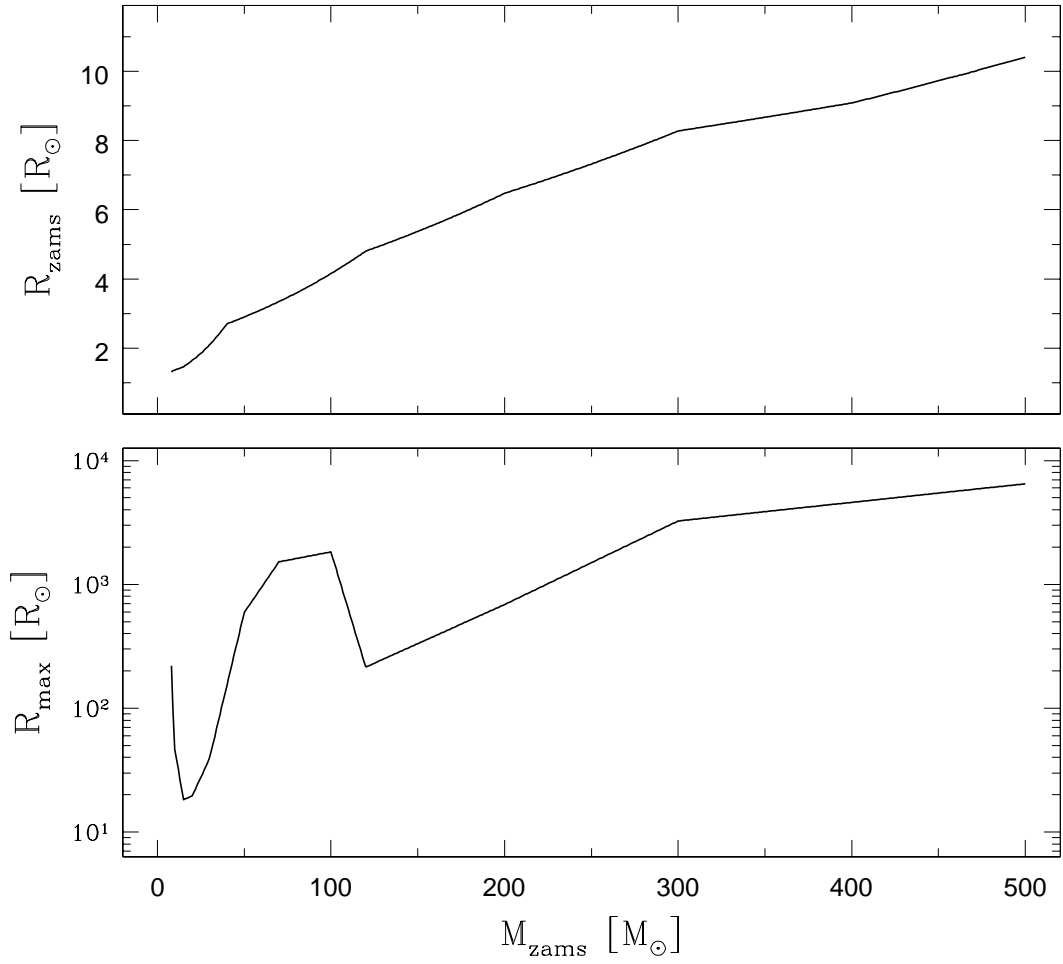}
\caption{Stellar radii for Population III stars; top panel shows initial
(ZAMS) radii, while bottom panel shows maximum radii. For details see
\S\,2.}
\label{radmax}
\end{figure}

The initial stellar radii are taken from the models above and from
Schaerer (2002).  By the end of the main sequence evolution, stellar
radii increase by a factor $\sim 2-3$ and, on the giant branch, their
radii can increase several orders of magnitude.  The expansion of a
star is very important since it is this expansion that determines the
onset of Roche lobe overflow (RLOF). We use models of Marigo et al.\ (2001)
and Baraffe et al.\ (2001) to obtain star effective temperature and
luminosity at the end of its evolution (maximum radial expansion), and
calculate corresponding radius for a given initial star mass. Initial,  
zero-age main sequence (ZAMS), radii and maximum radii for stars of
different initial mass are shown in Figure~\ref{radmax}.

Moment of inertia of a star is calculated from
\begin{equation}
I=k\ M\ R^2
\label{eq1}
\end{equation}
where (dimensionless) coefficients $k$ change with the evolution of star's
internal structure and are different for the envelope and the core. For MS
stars we use single coefficient $k_{\rm ms}=0.075$ (polytropic model for
massive stars; Lai, Rasio \& Shapiro 1993) since there is no well
defined core/envelope structure yet. For giants we adopt $k_{\rm env}=0.1$
(detailed giant models of Hurley, Pols \& Tout 2000) and
$k_{\rm cor}=k_{\rm He}$, and values for $k_{\rm He}$ are obtained from
the detailed models of Population III stars (given below). For exposed
helium cores we use $k_{\rm He}$ for the entire star.

Helium core mass is estimated from
\begin{equation}
 M_{\rm He} = \left\{ \begin{array}{ll}
0.4 M_{\rm zams} - 3 \msun & M_{\rm zams}<104 \msun \\
0.542 M_{\rm zams} - 17.8 \msun & M_{\rm zams} \geq 104 \msun \\
\end{array}
\right.
\label{eq2}
\end{equation}
where $M_{\rm zams}$ is an initial (ZAMS) mass of a star in $\msun$.

Helium core radius is taken from:
\begin{equation}
R_{\rm He} = \left\{ \begin{array}{ll}
-548000 \ M_{\rm zams}^2 + 4.92 \times 10^8 M_{\rm zams} +
1.11 \times 10^8 {\rm \, cm} & M_{\rm zams}<450 \msun \\
1.27 \times 10^{8} M_{\rm zams} + 0.53 \times 10^{11} {\rm \, cm} &
M_{\rm zams} \geq 450 \msun.  \\
\end{array}
\right.
\label{eq3}
\end{equation}
                                                                                                                              
Coefficients $k_{\rm He}$ are taken from Heger \& Woosley (2006), and
can be approximated by
\begin{equation}
k_{\rm He} = \left\{ \begin{array}{ll}
3.19 \times 10^{-7} M_{\rm zams}^2 - 0.000341 M_{\rm zams} + 0.3 &
M_{\rm zams} < 500 \msun \\
0.209 & M_{\rm zams} \geq 500 \msun. \\
\end{array}
\right.
\label{eq4}
\end{equation}

Moment of inertia $[g\ cm^2]$ of the very inner ($M_{\rm 7in}= 7 \msun$)
part of
a star core at the H depletion in the core (end of MS and He core formation)
is
estimated from
\begin{equation}
\log(I_{\rm 7in,He}) = \left\{ \begin{array}{ll}
53.82 + 0.192\ M_{\rm zams}
& M_{\rm zams}< 6.9 \msun \\
55.37 - 0.05\ M_{\rm zams} + 0.0016\ M_{\rm zams}^2 - 0.0000155\ M_{\rm
zams}^3 & 6.9 \geq M_{\rm zams}< 45 \msun \\
54.89 + 0.00148\ M_{\rm zams} - 0.0000017\ M_{\rm zams}^2
& 45 \geq M_{\rm zams}< 500 \msun \\
55.2
& M_{\rm zams} \geq 500 \msun, \\
\end{array}
\right.
\label{eq5}
\end{equation}
while the final value (at Fe core formation and just prior the collapse) is
estimated from
\begin{equation}
\log(I_{\rm 7in,Fe}) = \left\{ \begin{array}{ll}
57.25                                                & M_{\rm zams}< 10
\msun \\
68.73 - 1.79\ M_{\rm zams} + 0.0597\ M_{\rm zams}^2  & 10 \geq M_{\rm zams}<
15 \msun \\
59.79 - 0.36\ M_{\rm zams} + 0.00416\ M_{\rm zams}^2 & 15 \geq M_{\rm zams}<
43 \msun \\
52.0                                                 & M_{\rm zams} \geq 43
\msun. \\
\end{array}
\right.
\label{eq6}
\end{equation}
The above formulas describe the results of numerical
simulations with an accuracy better than a few percent.

\subsection{Remnant Masses/Supernova Explosions}

\begin{figure}
\includegraphics[width=0.9\columnwidth,angle=0]{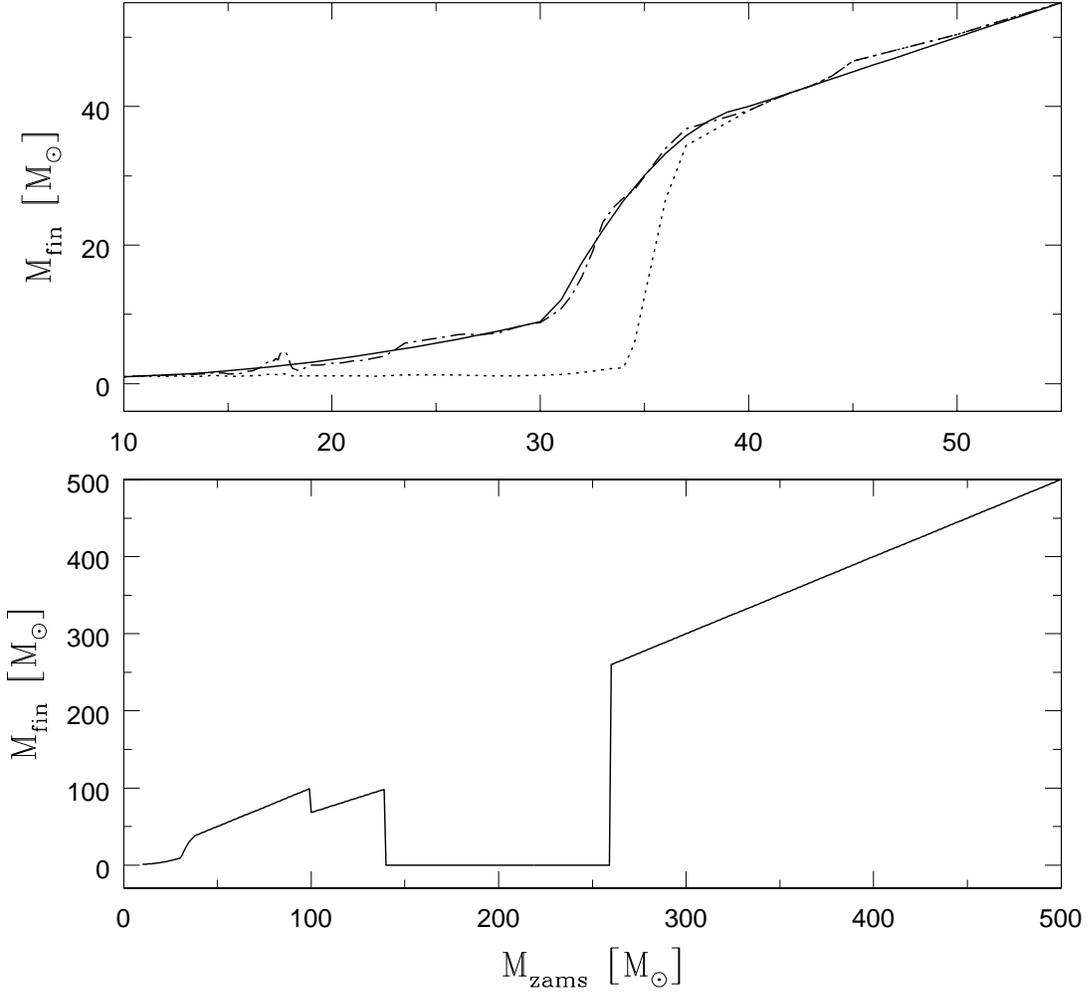}
\caption{Initial-final mass relation for Population III stars.
Bottom panel: our best fit to the Fryer (2006) models (see eq.~\ref{BHfin}).
Top panel: comparison with the Fryer (2006) models. Initial (before
fallback)
and final remnant masses are shown with dotted and dashed lines,
respectively.
Solid line indicates the fit used in this study. Note that for stars below
$\sim 35$\,M$_\odot$, the bulk of the remnant mass arises from fallback.
Note
also that because the structure of the star changes radically between 15 and
20\,M$_\odot$, so does the remnant mass.}
\label{mfin}
\end{figure}

Whether or not a given collapsing star produces a supernova explosion
is still a matter of intense study, but it has been known for quite
some time that if one can neglect mass loss, low mass stars are easier
to explode than high mass stars.  Using his 2-dimensional
core-collapse simulations, Fryer (1999) argued that low mass stars would
form neutron stars in a supernova explosion, but as the mass
increased, the explosion would become weaker and the fallback of
ejecta would collapse the compact remnant to a black hole.  Even more
massive stars would not produce a supernova explosion at all.
Assuming supernovae are driven by energy stored in the convective
region between the surface of the proto-neutron star and the accretion
shock, Fryer (2006a) developed a semi-analytic means to estimate the
explosion energy and final remnant mass of the collapsing star as
a function of the stellar structure. Figure~\ref{mfin} shows
the remnant mass as a function of initial star mass.
Fitting this semi-analytic estimate, and adding on the evolution of
more massive stars, we develop some simple expressions of the fate of a
Population III single star:
\begin{equation}
 M_{\rm fin} = \left\{ \begin{array}{lll}
0.015 M^2_{\rm zams} - 0.199 M_{\rm zams} + 1.491 \msun & 10 \leq M_{\rm
zams}<30.7 \msun & {\rm NS/BH} \\
-0.276 M^2_{\rm zams} + 22.69 M_{\rm zams} - 426.4 \msun & 30.7 \leq M_{\rm
zams}<40 \msun & {\rm BH} \\
M_{\rm zams} & 40 \leq M_{\rm zams}<100 \msun & {\rm BH} \\
M_{\rm zams}-f_{\rm loss} M_{\rm env} & 100 \leq M_{\rm zams}<140 \msun &
{\rm BH} \\
0 & 140 \leq M_{\rm zams}<260 \msun & {\rm no\ remnant} \\
M_{\rm zams} & M_{\rm zams} \geq 260 \msun & {\rm BH} \\
\end{array}
\right.
\label{BHfin}
\end{equation}
where we adopt $f_{\rm loss}=0.5$ for the fraction of the envelope lost by
a star due to pulsational pair-instability. We also adopt $M_{\rm max}^{\rm
NS}=3 \msun$, the limiting mass below which we call remnant a neutron star,
and over which BHs are formed. This translates into $M_{\rm zams} \sim 19
\msun$ for a minimum initial mass to form a BH.
Stars with mass below  $40 \msun$ lose part of their mass during the core
collapse/supernova event. For higher masses, the collapse occurs with
virtually
no mass loss, and the entire pre-supernova star mass ends up in the
final remnant. There are two exceptions for high mass stars. In the range
$100 \leq M_{\rm zams}<140 \msun$ part of star's mass is removed through
pulsational pair-instability and for $140 \leq M_{\rm zams}<260 \msun$ the
entire star is disrupted in pair instability supernova, with no remnant
left.

Most of the massive stars do not lose much mass in the final evolutionary
stages. However, if mass is lost, either through pulsational
pair-instability
or during core collapse and subsequent supernova explosion, the binary
system
may become eccentric. We assume spherical mass loss with no natal kick. We
verify that the system is bound and then solve the Kepler equations
for a new orbit.

\subsection{Binary Interactions}

Binary evolution may alter the final remnant mass of a given stellar
component.
During RLOF phases binary components may change their mass.
If a donor star is a main sequence star, we assume that such configuration
leads
to the component merger and we terminate binary evolution. If a donor star
is a giant,
we assume that it looses its entire H-rich envelope in the ensuing RLOF.
If a companion star accretes some of the donor material, it may either
become
rejuvenated (main sequence donors) or it simply increases its mass.
Rejuvenation is performed with the increase of $M_{\rm zams}$ and
recalculation
of all star properties. When a giant star increases its mass, only its
envelope
mass is increased while the core remains unaffected.

The above mass transfer affects final masses of compact objects. If a
star is rejuvenated, we use its new mass to calculate the final
remnant mass (i.e., in eq.~\ref{BHfin} new increased mass is used as
$M_{\rm zams}$).  In the case of mass loss, we assume that RLOF
removes the entire H-rich envelope, and then pre-supernova mass is
equal to the Helium core mass ($M_{\rm He}$).  The final mass is then equal
to either $M_{\rm He}$ or $M_{\rm fin}$, whichever number is smaller.

The binary orbit is allowed to change in the RLOF phases. Once a
star overfills its Roche lobe, we check for dynamical stability of
ensuing RLOF.  If the mass ratio of the donor to the accretor is larger
than $q_{\rm crit}=2$, we assume evolution into common envelope (CE)
phase.  Otherwise, non-conservative, but dynamically stable, mass transfer
(MT) is adopted. In both cases, the donor loses its entire envelope and
becomes an exposed Helium core.

For the CE phase we use standard energy balance (Webbink 1984) and
calculate the change of an orbit using:
\begin{equation}
 \alpha_{\rm ce} \left( {G M_{\rm don,f} M_{\rm acc} \over 2 a_{\rm f}}
   - {G M_{\rm don,i} M_{\rm acc} \over 2 a_{\rm i}} \right) =
{G M_{\rm don,i} M_{\rm don,env} \over \lambda R_{\rm don,lob}}
 \label{ce}
\end{equation}
where $M_{\rm don,env}$ is the mass of the donor envelope ejected from
the binary, $R_{\rm don,lob}$ is the Roche lobe radius of the donor at
the onset of RLOF, and the indices ${\rm i,\ f}$ denote the initial
and final values, respectively. In our calculations, we combine
$\alpha_{\rm ce}$ (CE efficiency) and $\lambda$ (the central
concentration of the donor) into one CE parameter, and we assume that
$\alpha_{\rm ce}\times\lambda = 1.0$.  In the case of the CE phase it
is assumed that no accretion on the companion takes place.

During dynamically stable MT episodes, part of the mass lost by the donor
($f_{\rm a}$) is accreted onto the companion, and the rest is lost from
the system with a specific angular momentum equal to ${2 \pi j_{\rm mt} a^2
/ P}$
(Podsiadlowski, Joss \& Hsu 1992), where $P$ denotes the orbital period. The
change
of orbit is calculated from
\begin{equation}
 {a_{\rm f} \over a_{\rm i}} =
 {M_{\rm don,f}+M_{\rm acc,f} \over M_{\rm don,i}+M_{\rm acc,i}}
 \left({M_{\rm don,f} \over M_{\rm don,i}}\right)^{c_1}
 \left({M_{\rm acc,f} \over M_{\rm acc,i}}\right)^{c_2},
\label{pod2}
\end{equation}
where
 $$c_1 \equiv 2 j_{\rm mt} (1-f_{\rm a})-2,$$
 $$c_2 \equiv - {2 j_{\rm mt} \over f_{\rm a}} (1-f_{\rm a})-2,$$
 $$M_{\rm acc,f}=M_{\rm acc,i}+f_{\rm a}(M_{\rm don,i}-M_{\rm don,f}).$$
We assume that half of the mass lost from the donor is also lost from the
system ($1-f_{\rm a}=0.5$) with specific angular momentum $j_{\rm mt}=1$,
while the remaining half is accreted onto the companion.

\subsection{Spin Evolution}

\begin{figure}
\includegraphics[width=0.9\columnwidth,angle=0]{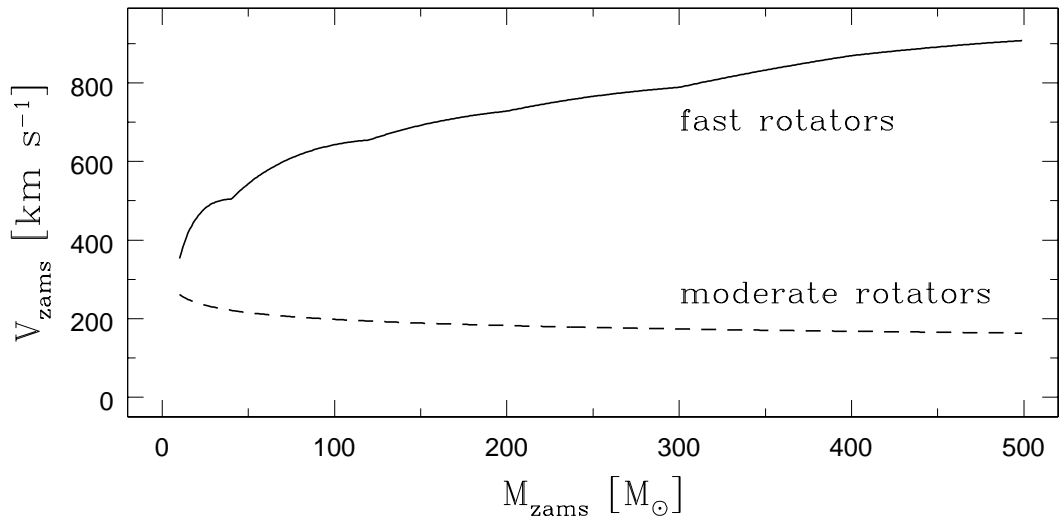}
\caption{
Initial rotational velocities adopted for Population III stars.
We use either 30\% of the Keplerian velocity (fast rotators), or we use the
estimates derived for Population I stars (moderate rotators). 
}
\label{vel}
\end{figure}

Since the initial rotational velocities of Population III stars are unknown,
we use two different assumptions for the initial spins.  All the stars are
initialized with rotational velocities  
\begin{equation}
V_{\rm zams} = \left\{
\begin{array}{ll}
13.4 \times M_{\rm zams}^{-0.12}/(0.0389+M_{\rm zams}^{-7.95}) & {\rm
moderate\ rotators} \\
0.3 \times V_{\rm break} & {\rm fast\ rotators} \\
\end{array}
\right.
\label{velzams}
\end{equation}
where $M_{\rm zams}$ is expressed in $\msun$, and $V_{\rm
break}=\sqrt{G M_{\rm zams}/R_{\rm zams}}$ is  the breakup (Keplerian)
velocity at the star equator.  In other words we have a model with fast
rotating stars: 30\% of the breakup velocity; and one in which we
adopt a distribution of rotational velocities equal to those used in
Population I stars. Population I initial velocities were taken
from a fit presented in Belczynski et al. (2006; see \S\,2.3.3), which
we extended to higher masses.  In Figure~\ref{vel} we show our
assumptions on the initial velocities. We note that Population III
stars (if initialized with $0.333 \times V_{\rm break}$) rotate much
faster ($\sim 400-900 {\rm km\ s}^{-1}$) than Population I stars
($\sim 200 {\rm km\ s}^{-1}$).

We then follow the evolution of the spin for each binary component.
The spin angular momentum of a star can be written as
\begin{equation}
J_{\rm spin} = \left\{
\begin{array}{ll}
k_{\rm ms} M R^2 \omega & {\rm MS\ star} \\
k_{\rm env} M_{\rm env} R^2 \omega_{\rm env} + k_{\rm He} M_{\rm He}
R_{\rm He}^2 \omega_{\rm He} & {\rm giants} \\
k_{\rm He} M_{\rm He} R_{\rm He}^2 \omega & {\rm He\ stars} \\
\end{array}
\right.
\label{jspin}
\end{equation}
where $M,\ R$ denote the mass and radius of an entire star, $M_{\rm He}$ is the He  
core mass and $M_{\rm env}=M-M_{\rm He}$ denotes the envelope mass, while
$R_{\rm
He}$ is the core mass. Spin of the entire star is marked with $\omega$, while
for the giants we denote the spin of He core and H envelope with $\omega_{\rm He}$
and
$\omega_{\rm env}$, respectively.

{\em Stage 1}\ \
During the main sequence, when no clear core-envelope structure exists,
we
assume uniform rotation of an entire star. Expansion on the main sequence (MS)
slows down a star. Final angular velocity of a star on terminal MS is
\begin{equation}
\omega_{\rm tms}=(R_{\rm zams}/R_{\rm tms})^2\ \omega_{\rm zams}
\label{rot01}
\end{equation}
where $\omega_{\rm zams}$ is the angular velocity at ZAMS, and the radius at
terminal MS
is approximated with $R_{\rm tms}=2.5\ R_{\rm zams}$. Initial angular
velocity
 of the  entire star is thus reduced by a factor $\sim 6$.  

{\em Stage 2}\ \
Just after star leaves main sequence, the clear core-envelope structure 
appears. Initially core and envelope angular velocities are the same, and
are obtained from the conservation of spin angular momentum while 
the star changes from  a uniformly rotating MS star to a giant 
with core and envelope characterized by different moments of inertia
\begin{equation}
\omega_{\rm bgb}= {k_{\rm ms}\ M\ R_{\rm tms}^2 \over
      k_{\rm He}\ M_{\rm He}\ R_{\rm He}^2 + k_{\rm env}\ M_{\rm env}\
R_{\rm tms}^2}
     \ \omega_{\rm tms}
\label{rot02}
\end{equation}
and then we set
\begin{equation}
\omega_{\rm env,bgb}=\omega_{\rm He,bgb}=\omega_{\rm bgb}.
\label{rot03}
\end{equation}
This leads to a mild spin up of the core and envelope by factors of
$\sim 1-1.5$ depending on its initial mass.

{\em Stage 3}\ \
The rotation of the helium core will likely decouple from the envelope (end
of main sequence/base of giant branch).
However, it is not known how effective such decoupling might be. Therefore,
during giant stage we follow the spins of the core and the envelope separately,
allowing for partial core-envelope coupling.
We introduce a simple spin He core--H envelope coupling to
obtain  
\begin{equation}
\begin{array}{l}
J_{\rm He,f}=X_{\rm cou1}\ J_{\rm He,i}\\
J_{\rm env,f}=J_{\rm env,i}+(1-X_{\rm cou1})\ J_{\rm He,i}
\end{array}
\label{rot04}
\end{equation}
where the indices i, f mark the angular momentum before and after the coupling,
respectively. For simplicity we perform coupling in one step at the time
just
prior core collapse (end of star evolution) or prior to envelope loss for
donors entering RLOF. The parameter $X_{\rm cou1}$ sets the strength of the
coupling, and we calculate models with different values: 0.01 (strong
coupling),
0.1 (weak), and 1.0 (none; core and envelope spins fully decoupled).
As a result, He core is spun down, while envelope is spun up.
The spin of a He core after coupling is
\begin{equation}
\omega_{\rm He,cou1}=X_{\rm cou1}\ \omega_{\rm He,bgb}
\label{rot05}
\end{equation}
and it decreases by a factor of $1-100$ depending on the adopted value of
$X_{\rm cou1}$.
The new envelope spin (readily obtained from eq.~\ref{rot04}) is not allowed
to be larger than the new core spin; i.e., the core slows down only to the 
point at which the spins of the core and envelope are equal.
Effectively, the final core spin may be in some cases larger than predicted
by eq.~\ref{rot05} and is obtained from
\begin{equation}
\omega_{\rm He,cou1}={J_{\rm He,i}+J_{\rm env,i} \over
k_{\rm He}\ M_{\rm He}\ R_{\rm He}^2 + k_{\rm env}\ M_{\rm env}\ R^2}.
\label{rot06}
\end{equation}
Also, if rotation of the  core is the same or slower than that of its
envelope, the coupling is not performed. Such situations may occur in close
binaries during the synchronization of the star.

{\em Stage 4}\ \ Late in the evolution, the inner region of the He core
condenses as it forms its iron core.  We calculate the change of rotation
in the inner $M_{\rm 7in}=7 \msun$ conserving its angular momentum,
i.e.,
\begin{equation}
\omega_{\rm 7in,Fe}= {I_{\rm 7in,He} \over I_{\rm 7in,Fe}}\ \omega_{\rm
He,cou1}
\label{rot07}
\end{equation}
where $I_{\rm 7in,He},\ I_{\rm 7in,Fe}$ denote moment of inertia of
the inner $7 \msun$ at the time of He core formation, and at the time
of iron core formation, respectively.  The inner region of the star
compresses (see a drastic change of the moment of inertia;
eq.~\ref{eq5}-~\ref{eq6}) and its angular velocity ($\omega_{\rm
7in,Fe}$) increases dramatically over its initial value ($\omega_{\rm
He,cou1}$).  The spin up factors are of the order of $\sim 3-100$ for
stars with initial masses of $19<M_{\rm zams}<29 \msun$, and $\sim
100-1500$ for masses $29<M_{\rm zams}<500 \msun$.

{\em Stage 5}\ \ 
We also perform a partial coupling of the inner $7 \msun$ and the
material surrounding it (the outer parts of He core; outside of inner
$7 \msun$) just prior to stellar collapse and remnant (and/or GRB)
formation. Following from the conservation of spin angular momentum
\begin{equation}
\begin{array}{l}
J_{\rm 7in,f}=X_{\rm cou2}\ J_{\rm 7in,i}\\
J_{\rm shell,f}=J_{\rm shell,i}+(1-X_{\rm cou2})\ J_{\rm 7in,i}
\end{array}
\label{rot08}
\end{equation}
we can obtain the final (prior the collapse) spin of the inner $7 \msun$
\begin{equation}
\omega_{\rm 7in,col}=X_{\rm cou2}\ \omega_{\rm 7in,Fe}
\label{rot09}
\end{equation}
where $J_{\rm shell},\ J_{\rm 7in}$ denote angular momentum of He shell
and inner $7 \msun$, respectively.
We use $X_{\rm cou2}$ to set the strength of coupling, and we calculate
models with different values: 0.01 (strong coupling), 0.1 (moderate
coupling)
and 1.0 (no coupling),
adopted from Woosley \& Heger (2006; see Table 1 for models with magnetic
field coupling included and excluded). As a result, the inner parts are
spun down by factors $\sim 1-10$. Magnetic torques were suggested as a
mechanism responsible for such coupling (Spruit 2002).
Note that, for massive stars, the iron core will be rotating faster (see
eq.~\ref{rot07}
and eqs.~\ref{eq5},~\ref{eq6}) than its surrounding He shell. We assume that
this is also the case if a star is subject to synchronization in a close
binary. This is justified by the fact that while synchronization may work
on a helium core for a prolonged time, once the iron core forms, it almost
immediately collapses. We will relax this assumption in our parameter study.  

{\em Stage 6}\ \ Stages 1-5 cover spin evolution of a single star or
non-interacting binary component. However, tidal interaction of stars 
in a short-period binary system may lead to an additional change of
stellar spins.  The strength of tidal synchronization/circularization
for massive stars is not well constrained due to lack of
observations. Therefore, we calculate two extreme models, one with and
one without tidal interactions allowed.  For the model with no tides the
spin evolution of each binary component ends at stage 5. In the
following we describe additional change of star spin due to the effect
of tidal interactions.

Once a star fills a significant part of its Roche lobe we assume that tidal
forces are strong enough to circularize the binary orbit and synchronize a
given
star. We use the criterion proposed by Portegies Zwart \& Verbunt
(1996)\footnote{The criterion was extended to include eccentric
binaries; factor (1-e) was added to check for onset of tides at the
closest (periastron) component encounter.} to invoke
circularization/synchronization
\begin{equation}
R_1 > 0.2 a (1-e)
\label{tid1}
\end{equation}  
where $a$ is semi-major axis of a binary orbit, $e$ its eccentricity and
$R_1$
is a radius of a binary component that is being synchronized.
Orbital parameters (semi-major axis, spin of the synchronized star) change
with conservation of the total angular momentum in the binary
\begin{equation}
J_{\rm 1,spin}+J_{\rm 2,spin}+J_{\rm orb}=const.
\label{tid2}
\end{equation}
where the individual stellar components are marked with indices $i=1,\
2$, orbital angular momentum is calculated from $J_{\rm orb} = M_1 M_2
\sqrt{a G (M_1+M_2) (1-e^2)}/(M_1+M_2)$, while component spin angular
momenta are taken from eq.~\ref{jspin} with current spin velocities
(as calculated through stages 1-5).  We solve eq.~\ref{tid2} for the
new semi-major axis $a$ requiring that the new orbit is circular
($e=0$) and the star synchronized. Synchronization is obtained with
the condition
\begin{equation}
\begin{array}{ll}
\omega_{1}=\omega_{\rm tide} & {\rm MS\ and\ He\ stars} \\
\omega_{1,env}=\omega_{1,He}=\omega_{\rm tide} & {\rm giants} \\
\end{array}
\label{tid3}
\end{equation}
where $\omega_{\rm tide} = \sqrt{G (M_1+M_2)} a^{-1.5}$ denotes the
mean angular orbital velocity of the new synchronized binary, and
index $1$ marks the component that is being synchronized. The spin of
the other component ($2$) is not affected.  Note that we allow tides
to work on an entire star (i.e., synchronization of both: envelope and
core in case of giants). Stars at different evolutionary stages (MS
stars, giants, naked He stars) are subject to synchronization provided
that the condition ~\ref{tid1} is fulfilled.  Depending on the orbital
separation of the system and the spin of the component that is being
synchronized, tides may either increase\footnote{It is checked that a
star is not spun up over breakup velocity.} or decrease a star's
angular velocity. We use population synthesis to determine the change
of stellar spins due to tidal interactions.

\subsection{Final Stages of Binary Evolution}

If both the stellar components of the binary collapse to form compact
objects (with or without accompanying GRB) without disrupting the
binary, it will form a double compact object binary; most frequently a
double black hole system. Such system will lose angular momentum
through an emission of gravitational radiation, and will eventually
merge, forming a single BH.  This coalescence can produce a strong 
burst of gravitational radiation (Kulczycki et al. 2006).

The binary evolution may be additionally terminated either by the
component mergers or binary disruption during supernova explosions.
If a star on the main sequence initiates RLOF we always assume it leads to
a merger, and single star formation. However, detailed evolutionary
calculations of RLOF have yet to be performed for metal free stars to
guide future population synthesis simulations. Additionally, we check
for the component coalescence after any RLOF phase. If any stellar
component radius is larger than its Roche lobe radius, a new single
object is formed and we stop the evolution. It is noted that some of
the merger products may be still potential GRB progenitors, but are
not considered here, e.g. mergers of two exposed He-cores (Fryer,
Hartmann, \& Woosley 1999) or the merger of a compact remnant with a
He-core companion (Fryer \& Woosley 1998). We will analyze merger
products as Population III GRB progenitors in a separate paper.

The binary may be disrupted because of the mass ejected in either the
supernova (or GRB) explosion in the collapse of the stellar components
or through mass loss caused by pulsations\footnote{Although in most
cases, the entire pre-supernova star eventually ends up in the remnant
(see eq.~\ref{BHfin}).}.  The binary can also be disrupted if the
stellar component undergoes a pair-instability explosion, leaving
behind no remnant whatsoever.  We assume that there are no natal kicks
involved in the collapse/supernova of Population III stars. If
Population III stars are subject to natal kicks, many Population III
binaries are disrupted, terminating potential paths leading to a GRB
phenomenon.

\subsection{Initial Parameter Distributions/Standard Model}

The initial mass function (IMF) of Population III stars is only known from
numerical simulations (see O'Shea \& Norman 2006 for most recent study). 
It is found to be top heavy and could be bimodal
with a high mass peak around $100 M_\odot$ (Bromm et al. 1999, 2002;
Omukai \& Palla 2003) . The upper limit on the mass range is roughly
$500\, M_\odot$ (Bromm \& Loeb 2004). The lower limit is  less certain and
has been estimated to be $30\,M_\odot$ (Tan \& McKee 2004),
or $100\,M_\odot$ (Abel, Bryan \& Norman 2002; Bromm \& Larson 2004).
The shape of the Population III mass function has been parametrized in
different ways: a broken power law with $M_{break}=100 M_\odot$, low mass
slope of 0 and a high mass slope $2<\beta< 3$ (Belczynski, Bulik \& Rudak
2004), or a single power law with the Salpeter exponent (Bromm \& Loeb 2006).

\begin{figure}
\includegraphics[width=0.9\columnwidth,angle=0]{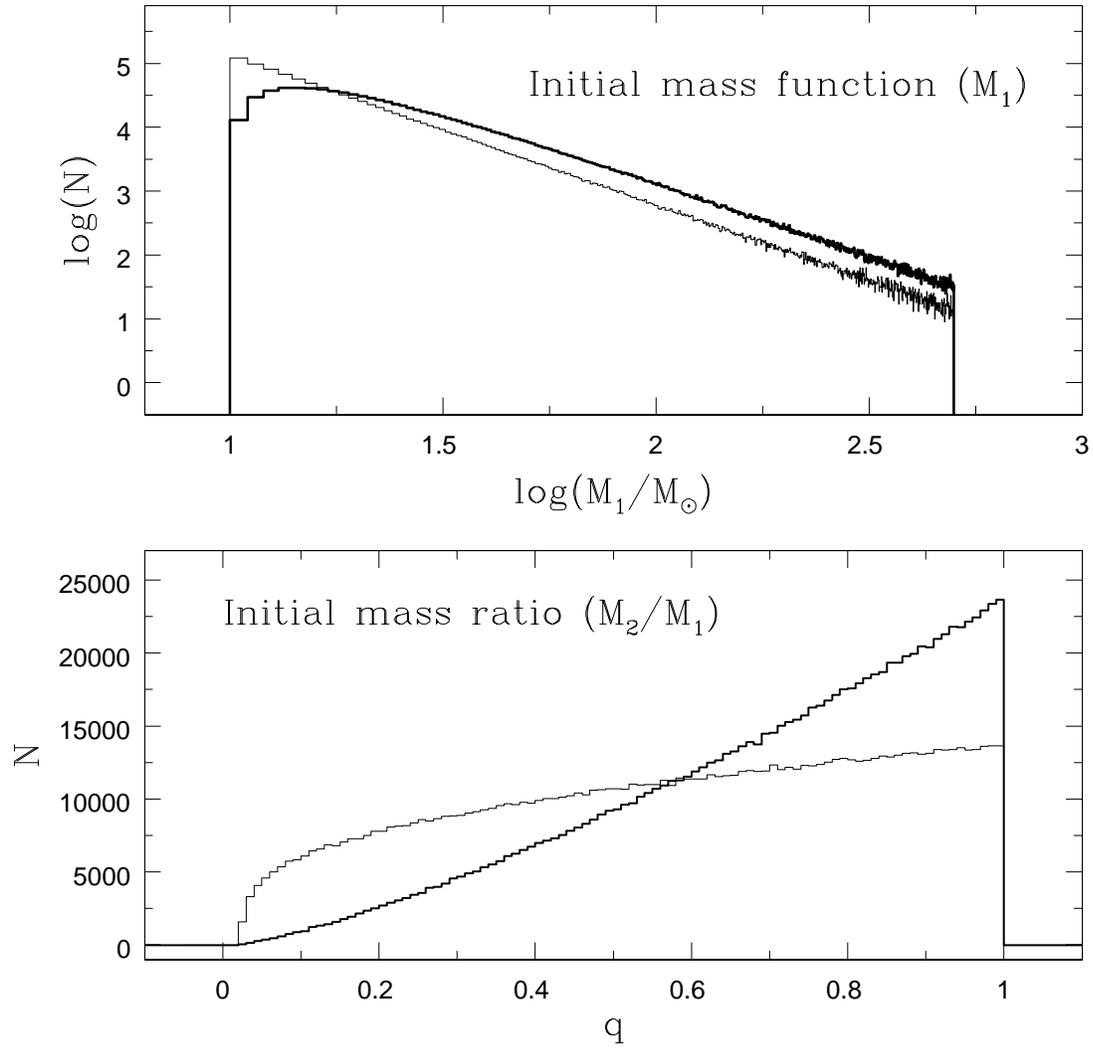}
\caption{
Initial mass function and initial mass ratio distribution for Population
III stars; standard model (thick lines) and alternative IMF model
(thin lines) are presented. For details see \S\,2.6.
}
\label{imf}
\end{figure}

In our standard model we choose a mass range $10-500 \msun$. We draw a
primary mass ($M_1$) from a power-law distribution $\Psi(M)\propto
M^{-\beta}$
with $\beta=2.35$ and within the adopted mass range. Next we draw a mass
ratio
($q=M_2/M_1$) from flat distribution in range $q=0-1$. If a secondary mass
($M_1$) is lower than the lower limit of the adopted mass range ($10 \msun$)
we repeat the procedure, drawing primary mass and mass ratio again, until the
secondary mass lies within the desired mass range. The resulting primary
mass IMF is flat at low masses ($M_1 \lesssim 20$) and then falls off as a
steepening power-law ($\sim M^{-1.8}-M^{-2.0}$). The resulting mass ratio
is skewed toward high-$q$ values. Both distributions are presented in
Figure~\ref{imf}. An average mass of a primary is $M_{\rm av,1}=44.8 \msun$,
and secondary $M_{\rm av,1}=27.4 \msun$. Therefore, average mass of a star
in a binary system is $M_{\rm av,bin}=36.1 \msun$, while a single star would
have an average mass $M_{\rm av,sin}=44.8 \msun$ (drawn from the same
distribution as a primary star). The effect of a change of shape and mass
range of IMF on the results is investigated in our parameter study.

The distribution of initial orbital separations is flat in logarithm like
it is usually adopted for the Population I stars. Systems are allowed to
form from the minimum separation, such that the stars are not in contact:
$a_{\rm min}=1.3\ R_{\rm 1,zams} + 1.3\ R_{\rm 2,zams}$, to maximum
separation
of $a_{\rm max}=10^6 \rsun$. Initially all systems are placed on circular
orbits. In each model calculation we evolve $N_{\rm bin}=10^6$ primordial
Population III binaries. We also adopt a binary fraction of $f_{\rm
bin}=0.1$
for Population III stars.  

In addition, in our standard model, we use 30\% of Keplerian velocity as
initial star rotation speed, we allow tidal interactions (with tides
working on an entire star), no natal kicks are applied at compact object
formation, maximum NS mass is $3 \msun$, CE efficiency is $\alpha_{\rm
ce}\times\lambda = 1.0$, non-conservative evolution is applied during
dynamically stable RLOF with $f_{\rm a}=0.5$ and $j_{\rm mt}=1$, coupling
between He core and H-rich envelope is applied with $X_{\rm cou1}=0.1$, and
also coupling between Fe core and He shell is applied with $X_{\rm
cou2}=0.1$.

\subsection{GRB Classification}

We require that a star to make a GRB needs to:  
{\em (i)} lose its H-rich envelope and become a naked He star,
{\em (ii)} the collapsing inner core must have a significant
         amount of angular momentum,
{\em (iii)} and that the collapsing core forms a black hole remnant.  
Following evolution of each binary we easily select the stars that
lose their envelopes and in the end form BHs.  We also follow the 
evolution of the angular momentum in the star, obtaining an expression 
for the inner 7\,M$_\odot$ core:
\begin{equation}
J_{\rm 7in}= I_{\rm 7in,Fe} \omega_{\rm 7in}
\end{equation}
where $I_{\rm 7in,Fe}$ is given by eq.~\ref{eq6} and $\omega_{\rm 7in}$
is calculated from the spin evolution of each star in the simulation
(see \S\,2.4). We also introduce the dimensionless specific angular
momentum
\begin{equation}
\tilde j_{\rm 7in} = J_{\rm 7in} c/G M_{\rm 7in}^2
\end{equation}
where $c$ is the speed of light, $G$ gravitational constant, and
$M_{\rm 7in}$ is the mass of the inner part of a collapsing core.

In order to form a disk around the newly formed black hole the
specific angular momentum of the material must be large enough to
prevent instant accretion. The specific angular momentum of a particle
orbiting on the marginally stable orbit for a non-rotating black hole
is $\tilde j=\sqrt{6} $. This condition (Podsiadlowski et al. 2004)
was mentioned by Bromm \& Loeb (2006), however, they did not verify if
it is satisfied for their GRB progenitors originating from Population
III binaries. Numerical simulations of a collapsing core (MacFadyen \&
Woosley 1999) showed that for the case of a $14 \msun$ core a disk
will form if the specific angular momentum before the collapse is in
the range of $0.42<\tilde j <2.8$. For the core rotating too fast
there was no significant accretion and a modest outflow took place.
For each collapsing core in our simulations we calculate the value of
its dimensionless specific angular momentum $\tilde j_{\rm 7in}$. The
value $\tilde j=1$ corresponds to the maximally rotating Kerr black
hole.  The requirement to make a GRB is that $\tilde j$ is
close/larger than unity, and we consider three different potential
ranges:
\begin{equation}
\begin{array}{ll}
0.42 < \tilde j_{\rm 7in} <2.8 & {\rm MacFadyen\ \&\ Woosley\ (1999)\ (A)}\\
   1 <   \tilde j_{\rm 7in} <10 & {\rm this\ work\ (B)}\\
\tilde j_{\rm 7in} > \sqrt{6} & {\rm Podsiadlowski\ et\ al.\ (2004)\ (C)}\\
\end{array}
\label{eq01}
\end{equation}
where model (B) comes from the requirement that
the newly formed black hole is spun fast enough to become
maximally rotating Kerr black hole,
and the accreting  material  has enough angular momentum
to form a disk. As we will see below the particular choice of a given
criterion A, B, or C  does not change the results significantly.

\subsection{SWIFT Detection Rates}

We assume that the SWIFT detection threshold is $f_{\rm min}=0.2 s^{-1}$
(Berger et al. 2005). The GRB luminosity function has been estimated by
Sethi \& Bhargavi (2001) who show that it can be well approximated
by a lognormal distribution
\begin{equation}
\Phi(L)= {e^{-\sigma^2/2}\over L_0 \sqrt{2\pi\sigma^2}}
\exp\left\{ -{[\ln(L/L_0)]^2\over 2\sigma^2  }\right\}
\end{equation}
centered on $L_0=2\times 10^{56}s^{-1}$, with a width of
$\sigma=2$.
We can now easily find the
GRB rate by integrating:
\begin{equation}
R_{SWIFT}={\eta_{mass} \eta_{beam} \eta_{S} \eta_i \over M_{av,bin}}
\int \Phi(L) dL \int_0^{z_{max}(L)} {SFR(z)\over 1+z} {dV\over dz} dz
\label{Rone}
\end{equation}
where $\eta_{beam}$ is the beaming correction, $\eta_{S}$ is the SWIFT
GRB detection efficiency, $\Psi(L)$ is the luminosity function, and
$SFR(z)$ is the Population III star formation rate. The GRB formation
efficiency $\eta_i$ is the number fraction of stars in Population III
binaries that make GRBs (see eq.~\ref{eff}).

The fraction of entire Population III mass that is contained in the
binaries is denoted as $\eta_{mass}$. We define binary fraction as
$f_{\rm bin}=N_{\rm bin}/(N_{\rm bin}+N_{\rm sin})$, where $N_{\rm bin}$
and $N_{\rm sin}$ denote the number of binary and single stars in a
given population, respectively. If so, then the mass fraction is
\begin{equation}
\eta_{mass} = {2 f_{\rm bin} M_{av,bin} \over 2 f_{\rm bin} M_{av,bin} +
              (1-f_{\rm bin}) M_{av,sin}}
\label{etamass}
\end{equation}
where $M_{av,bin}$ denotes the average star mass in a binary system, while
$M_{av,sin}$ is an average mass of a single star, both of which depend
on adopted IMF.
For the standard model IMF, we have $M_{av,bin}=36\,M_\odot$ and
$M_{av,sin}=45\,M_\odot$, and then for the adopted binary fraction $f_{\rm
bin}=0.1$ we obtain $\eta_{mass}=0.15$.

The limiting redshift is obtained by solving
\begin{equation}
d_l(z_{max})(1+z_{max})^{(\alpha-2)/2} = \sqrt{\displaystyle L\over 4\pi
f_{min}}
\end{equation}
where $d_l(z)$ is the luminosity distance, and $\alpha$ is the photon
index of the GRB spectrum. We assume that $\alpha=2$.
The Population III star formation rate is poorly known
therefore we will parametrize its shape by box function:
\begin{equation}
SFR(z)=S \Theta(z-z_{start})\Theta(z_{stop}-z)
\label{sfr}
\end{equation}
where $\Theta(z)$ is the Heaviside function.
With this choice equation~\ref{Rone} can be simplified to
\begin{equation}
R_{SWIFT}= 3 {\rm yr}^{-1}  \eta_i \left({\eta_{mass}\over 0.15} \right)
 \left({\eta_{beam}\over 0.02} \right)
 \left({\eta_{S}\over 0.1} \right)
 \left({S\over S_0 } \right)
 \left({36M_\odot\over M_{av}}\right)  
 \left({F(z_{start})-F(z_{stop}) \over 3.93}\right)
 \label{rate}
\end{equation}
where $S_0=10^{-2}M_\odot {\rm Mpc}^{-3}$ and   $F(z)$ is
\begin{equation}
F(z)=(10^{7} {\rm Mpc}^3)^{-1} \int \Phi(L) dL \int_0^{z_{max}(L)}
{\Theta(z-z')\over 1+z'} {dV\over dz} dz'\, .
\end{equation}
The typical values of the rightmost factor in equation~\ref{rate},
$\left(F(z_{start})-F(z_{stop})\right)$, lie in the range between one and
ten.
Thus the rate is directly proportional to $\eta_i$ -- the GRB production
efficiency, which is obtained with evolutionary calculations of Population
III binaries. If GRB production from Population III binaries is very efficient, 
one can
expect a few events detectable by SWIFT per year.

\section{Results}

In the following subsections we first present results for our standard
evolutionary model (see \S\,2.6), and then show the sensitivity of
results
on a number of adopted parameters. For the sake of presentation, we will
consider ``{\em potential GRB progenitors}'' defined as stars that are
stripped of their H-rich envelopes and are massive enough to form BH (but no
constrains are put on  their specific angular momentum) and ``{\em GRB
progenitors}''
for which we additionally apply angular momentum constraints (see
eq.~\ref{eq01}).

\subsection{Evolutionary Scenarios}

We consider a number of evolutionary channels to form potential GRB
progenitors in Population III binaries.  The major channels are
presented in Table~\ref{tab01}. GRB progenitors evolve through RLOF by
either dynamically stable MT or dynamically unstable CE phases, which
expose the He cores of the stellar component/components. A star
initiating a given event is indicated with a number: 1 stands for
primary (initially more massive star) and 2 for secondary binary
component.  By potential progenitor we consider any star that has been
stripped of its H-rich envelope that will form BH. Only $\sim 10$\% of
stars in Population III binaries end up as exposed He cores massive
enough to form BH.  In particular it is found that in some cases both
the primary and secondary components are stripped of their H-envelopes
and may produce a potential GRB.  The most efficient channel (grb01)
involves both MT and CE episode and $\sim 7$\% binaries. However, if
constraints on the angular momentum are imposed (see eq.~\ref{eq01})
only small fraction of binaries forms cores with high angular momentum
to produce GRB. Table~\ref{tab01} lists formation efficiencies along
each channel for different constraints on angular momentum. The
formation efficiency (per star in Population III binary) is defined as
\begin{equation}
\eta_i={N_{\rm grb} \over 2 N_{\rm bin}}
\label{eff}
\end{equation}
where $N_{\rm bin}=10^6$ is the total number of binaries evolved in a
given simulation, $N_{\rm grb}$ the number of stars that have produced
a GRB, and $i$ denotes angular momentum constrain used: in case of no
angular momentum constraints ($\eta_1$); MacFadyen \& Woosley 1999
($\eta_2$); this work ($\eta_3$); and Podsiadlowski et al. 2004
($\eta_4$).  It is found that if angular momentum constraints are
imposed, GRBs are found to be formed only through the grb01
channel. In particular, only a small fraction of secondaries formed
through channel grb01 have sufficiently high angular momentum to
produce a GRB. Formation efficiencies are very small, $\eta_2=0.0002,\
\eta_3=0.0058,\ \eta_4=0.0063$ due to very specific conditions
required to form a fast rotating core of a massive star in Population
III binary.  In the following we give a brief description of a sample
evolution of a progenitor system (channel grb01) leading to the
formation of a GRB.

We take an example of an initial binary consisting of a $M_1=40 \msun$
primary and $M_2=31 \msun$ secondary on an orbit with a semi-major axis of
$a=75 \rsun$  (orbital period $P_{\rm orb}=8.9$ day).  
The maximum radii are $161$ and $44 \rsun$ for the
primary and secondary components, respectively. Since their Roche
lobes are only $\sim 30$ and $27 \rsun$ the system is bound to evolve
through two RLOF episodes.
Stars are initiated with the uniform angular rotation velocities of
$\omega_1=23.1\ {\rm day}^{-1}$ and $\omega_2=28.8\ {\rm day}^{-1}$.  

As the primary expands onto the giant branch, it synchronizes as it
fills its Roche lobe; its envelope spins up while core slows down to and
angular velocity ($\omega_1$) equal to $2 \pi / P_{\rm orb} = 0.7\
{\rm day}^{-1}$. Since the core and envelope are spinning at the same
rate there is no coupling (angular momentum transport from core to
envelope) performed in this case.  
Change of the primary spin is done at the cost of orbital angular momentum
and the orbit decays to $a=72 \rsun$ ($P_{\rm orb}=8.5$ day). The  
first mass-transfer episode is dynamically stable (comparable component
masses) and the orbit tightens only moderately $a=42 \rsun$ ($P_{\rm
orb}=4.1$ day). The primary loses its envelope and becomes a naked He star
($M_1=13
\msun$), while the secondary accretes a fraction of the mass and is
rejuvenated ($M_2=45 \msun$). Now we check if the He star primary (radius
of $R_1=0.3 \rsun$) fills a significant part of its Roche lobe to undergo
further synchronization. Since this is not the case ($R_1 \ll a$), the He
star
spin remains unchanged ($\omega_1= 0.7\ {\rm day}^{-1}$).
Then the inner core compresses, and we spin it up to from $\omega_1= 0.7\ {\rm
day}^{-1}$ to $\omega_1= 594\ {\rm day}^{-1}$ to take into account the 3
order
of magnitude decrease of the moment of inertia of the inner parts of the He
core.  
Just before the collapse we perform the coupling, and the inner core loses
a fraction of its angular momentum ($X_{\rm cou2}=0.1$) to the surrounding He
material, and it slows down to $\omega_1= 59.4\ {\rm day}^{-1}$.  
In this mass range the star forms a BH with no mass loss ($M_1,BH=13 \msun$).
As the star collapses the specific angular momentum of the inner part of the
collapsing star is only $\tilde j \sim 0.02$, and the star does not
produce a GRB due to insufficient angular momentum. After SN the
orbit
remains bound and unaffected (no mass loss, no natal kick).

Then the secondary evolves off its main sequence ($\omega_2=3.5\ {\rm
day}^{-1}$), and while
climbing up the giant branch and filling its Roche lobe, it synchronizes to
$\omega_2= 1.8$ day$^{-1}$. Coupling between He core and H envelope is not
performed since both rotate with the same angular velocities.
The secondary initiates a second RLOF episode at
$a=37$ ($P_{\rm orb}=3.4$ day). Because the donor ($M_2=45 \msun$) is now
much
more massive than its BH companion ($M_{\rm 1,BH}=13 \msun$) the following
RLOF
is dynamically unstable and the system enters CE phase.  The orbit tightens
significantly to $a=1.2 \rsun$ ($P_{\rm orb}=0.029$ day), the secondary
loses its
envelope and becomes an exposed helium core ($M_2=14.8 \msun$). Since the
post-CE orbit is so small, the Helium star ($R_1=0.3 \rsun$) fills
a significant part
of its orbit ($R_1 > 0.2 a$) and synchronizes. After synchronization, the
star is rapidly spinning ($\omega_2=250\ {\rm day}^{-1}$) and the orbit
tightens slightly to $a=1.1 \rsun$ ($P_{\rm orb}=0.025$ day).
When the core condenses during oxygen and silicon burning, it is spun up to $\omega_1= 221100\ {\rm day}^{-1}$
through decrease of moment of inertia of the material forming it. Finally,
we
perform coupling of the Fe core with He shell ($X_{\rm cou2}=0.1$), spinning
the inner core down to $\omega_1= 22110\ {\rm day}^{-1}$.  
Fast rotation of the inner core results in a large specific angular
momentum of the secondary at the time of the collapse, $\tilde j_2 \sim 6$.
Depending on the angular momentum criterion, the secondary may (B, C) or may
not (A) produce a GRB. The secondary after the collapse forms BH with mass
$M_1,BH=14.8 \msun$. The close massive BH-BH binary forms with $a=1.1 \rsun$
($P_{\rm orb}=0.025$ day) and its components merge due to the emission of GR
in less than 0.1 Myr.

\subsection{Effects of Tides}

At every step the population synthesis code takes a new semi-major axis $a$
and subsequently new stellar spin $\omega_{\rm tide}$ is established.
The change of the core angular velocity due (only) to the tidal effects is
calculated as
\begin{equation}
d_{\rm cor,vel} = {\omega_{\rm 1,f} \over \omega_{\rm 1,i}}
\label{tid4}
\end{equation}
where $\omega_{\rm 1,i}$ denotes the initial (pre-synchronization) velocity
of He core, and $\omega_{\rm 1,f}=\omega_{\rm tide}$ is the
post-synchronization velocity.

\begin{figure}
\includegraphics[width=0.9\columnwidth,angle=0]{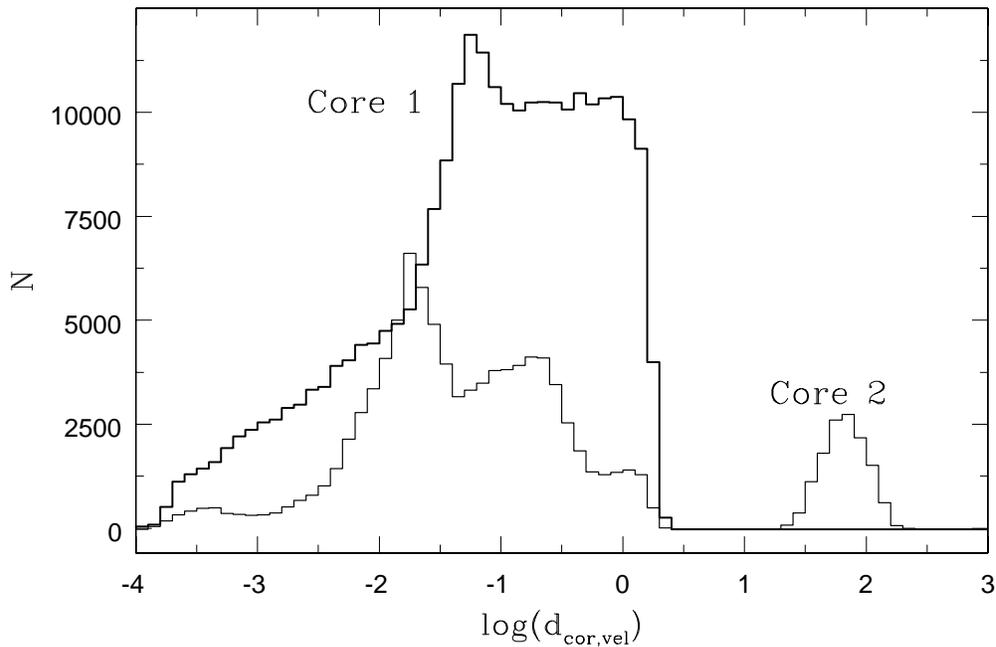}
\caption{
Effects of tides (synchronization) on core angular velocity. A
core of a binary component is generally slowed down (speed
changes by factor of $d_{\rm cor,vel} \sim 0.01-1$) if tidal
forces are allowed to work on entire star. Core speed change
for a primary (core 1) and a secondary star (core 2) are presented.
Note a subpopulation of secondary cores that are spun up by
factors of $d_{\rm cor,vel} \sim 30-100$. For details see \S\,2.4.
}
\label{tide}
\end{figure}

The effects of tides on core angular velocity are presented in
Figure~\ref{tide}.
For our reference model we have recorded change of the core rotational velocity
($d_{\rm cor,vel}$) for primaries and secondaries.  
Only $\sim 25\%$ primaries, and $\sim 5\%$ secondaries are subject to
synchronization, as systems are disrupted in SNa explosions, binary
components merge in RLOF events, or some systems are simply too wide to
interact.

In most cases the tides are effective when a star is already on the
giant branch, and has a large (as compared to the orbit size)
radius. By that time, the core has decoupled from the envelope; and
its spin is much larger than that of an expanding envelope. Generally,
once tides become effective, the mean orbital angular velocity is smaller
than the core angular velocity, and the subsequent synchronization slows
down the stellar core. For cores which are affected by tides during
the giant branch, the speed decreases by factors of $\sim 1-1000$
($d_{\rm cor,vel} \sim 1-0.001$).  However, we find a subpopulation of
secondary cores with a rather high tidal speed-up factors $d_{\rm
cor,vel} \sim 30-100$.  In some cases, a naked Helium star is subject
to synchronization.  This happens only for small orbits since the
radii of Helium stars are rather small ($< 2 \rsun$; see
eq.~\ref{eq3}). Such configuration is encountered in the late
evolutionary stages of a massive binary. After the primary has already
formed a compact object, the secondary may initiate a common envelope
phase, that leads to a drastic orbital tightening. As a result, a
short period binary with secondary He star emerges. If the orbit is
sufficiently small, tides are operating and the He star is
synchronized. Since the mean orbital angular velocity is then rather
large, the Helium star can spin up significantly.

\begin{figure}
\includegraphics[width=0.9\columnwidth,angle=0]{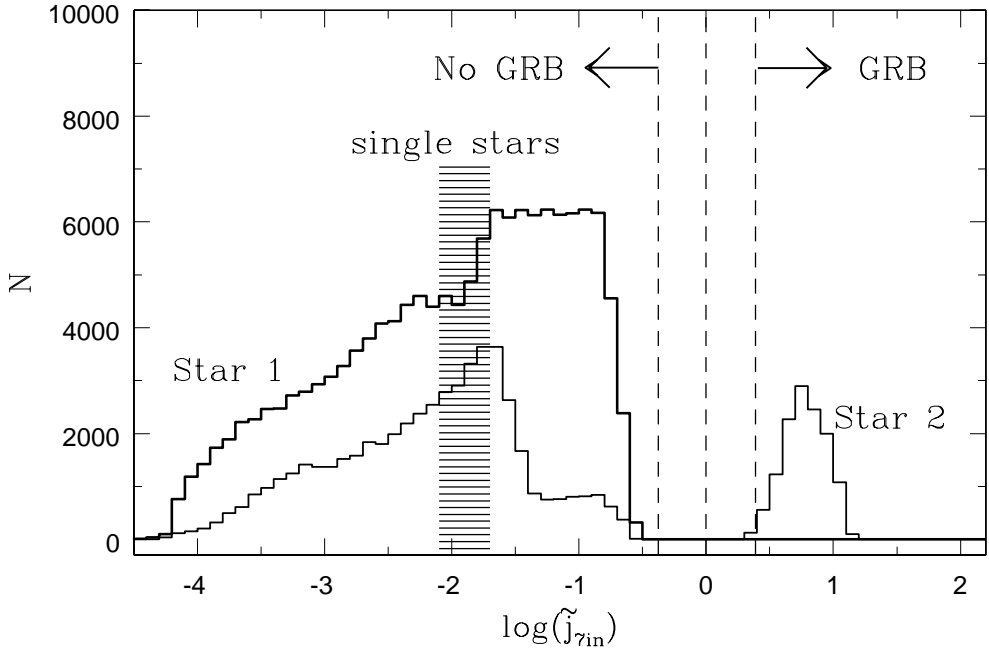}
\caption{
Specific angular momentum of potential GRB progenitors from
Population III binaries: primary is marked as Star 1, while
secondary as Star 2. Potential progenitors
are these stars that are stripped of their envelopes and are massive
enough to form BH (but no constrain is put on their specific angular
momentum). For comparison we plot the range (shaded
area) of specific angular momentum for Population III single
stars (these which in the end form black holes).
We also mark (dashed lines) the minimum specific angular
momentum required to produce GRB for three different criteria
used in this work (see eq.~\ref{eq01}). Note that only a
subpopulation of secondary stars with rather high specific
angular momentum ($\tilde j \gtrsim 1$) will produce GRB.}
\label{angm}
\end{figure}

In Figure~\ref{angm} we show the final specific angular momentum of the
inner core for our reference model. For single stars (that are in the mass
range to form black holes) we find $\tilde j_{\rm 7in} \sim 0.01-0.02$
which is a direct result of the adopted single stellar models (\S\,2.1)
and assumptions stated in \S\,2.4. Primary stars are characterized by
$\tilde j_{\rm 7in} \sim 10^{-4}-0.1$. The low values $\tilde j_{\rm
7in}<0.01$ are the result of tidal spin down on giant branch. The high
values $\tilde j_{\rm 7in}>0.02$ are a combination of the synchronization
without spin down (see Figure~\ref{tide}; not all of the cores are
significantly slowed down) that leads to a lack of angular momentum
transport since the core rotates with the same angular velocity as envelope.
Secondary stars show a bimodal distribution
of the core specific angular momentum. Low values $\tilde j_{\rm 7in}
\sim 10^{-3}-0.1$ are found for secondaries that were spun down during giant
branch, while high values $\tilde j_{\rm 7in} \sim 2-10$ are found for
secondaries that were spun up as a naked Helium stars.
Only for this last subpopulation is the specific angular
momentum high enough to warrant a potential GRB.

\subsection{Properties of GRB Progenitors}

\begin{figure}
\includegraphics[width=0.9\columnwidth,angle=0]{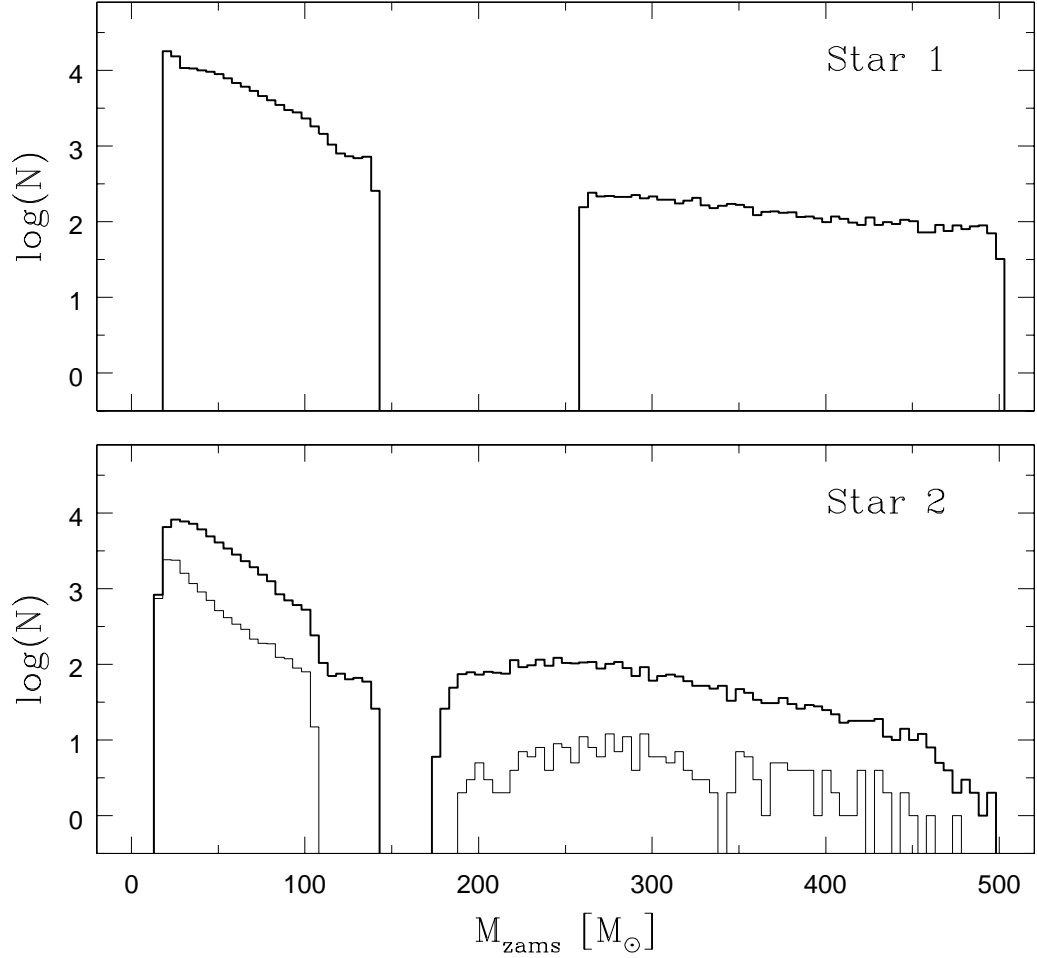}
\caption{
Distribution of initial (ZAMS) masses in potential (naked helium stars
massive enough to form BH) GRB binary progenitors. Potential progenitors
are these stars that are stripped of their envelopes and are massive
enough to form BH (but no constrain is put on their specific angular
momentum). Primary components are shown in the top panel,
while secondaries in the bottom panel.
Progenitors that produce cores with specific angular momentum
($\tilde j \gtrsim 1$) high enough to make GRB are shown
with the thin line.
}
\label{mzams}
\end{figure}

\begin{figure}
\includegraphics[width=0.9\columnwidth,angle=0]{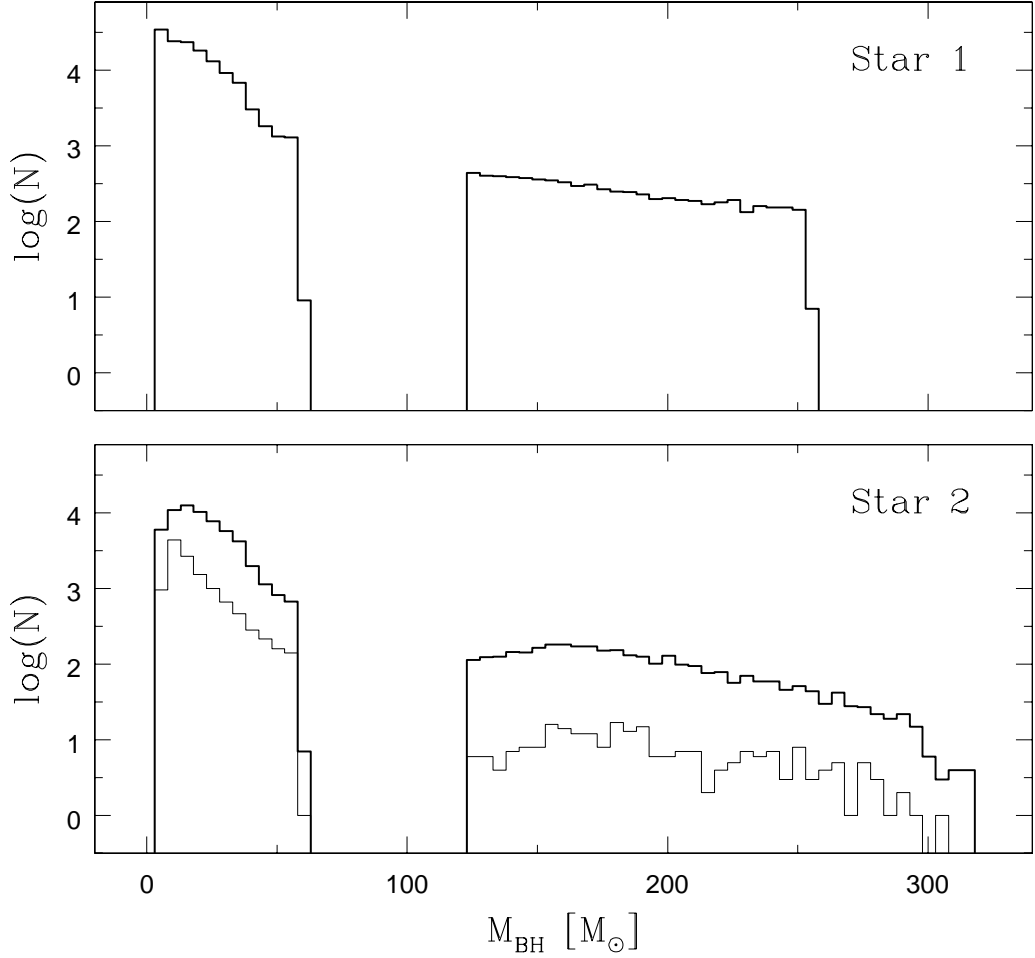}
\caption{
Distribution of final remnant (BH) masses in potential GRB binary
progenitors.
Potential progenitors
are these stars that are stripped of their envelopes and are massive
enough to form BH (but no constrain is put on their specific angular
momentum).
Remnants from a primary component are shown in the top panel,
while from secondaries in the bottom panel.
Progenitors with specific angular momentum ($\tilde j \gtrsim 1$)
high enough to produce GRB are shown with the thin line. Note that only
small fraction of secondaries can produce a GRB.
}
\label{mf}
\end{figure}

In Figure~\ref{mzams} and ~\ref{mf} we present the initial (ZAMS) masses of
potential GRB progenitors and the BH masses they form, respectively.  
The initial mass distribution spans almost the entire investigated mass range
$M_{\rm zams}=10-500 \msun$. However, primaries only over $M_{\rm zams}
\gtrsim 19 \msun$ may form BHs (and only such stars are included as
potential GRB
progenitors).  
There are no primaries in the initial mass range $140-260 \msun$ because such 
stars become 
pair instability supernovae. For secondaries the range is narrower
$\sim 140-180 \msun$ due to rejuvenation, and in particular stars over $\sim
180 \msun$ may gain enough mass ($\gtrsim 80 \msun$; half of the primary
envelope) to be shifted out of pair instability supernovae mass strip.  

Black holes are found within mass range $\sim 3-250$ and $\sim 3-320 \msun$
for primary and secondary components, respectively. The low-mass cut
reflects our assumption on the minimum BH mass, and the high-mass cut
corresponds to He core mass of the most massive star in the simulation;
$M_{\rm He}=250 \msun$ ($M_{\rm zams}=500 \msun$) for primaries, and $M_{\rm
He}=320 \msun$ ($M_{\rm zams}=630 \msun$) for rejuvenated secondaries.
Pair instability supernova prevent the formation of BHs in secondaries 
within the mass range between $\sim 60$ and $120 \msun$. Here, the
empty interval is the same for both components. The stripped primaries
(potential GRB progenitors included here) just outside of pair instability
range form BHs with masses 60 ($M_{\rm zams}=140$) and $120 \msun$
($M_{\rm zams}=260 \msun$). Secondaries with initial masses that fall
within pair
instability supernova mass range ($M_{\rm zams} \sim 180-260 \msun$) gain
enough mass and are rejuvenated to form BHs as stars with $M_{\rm zams} >
260 \msun$. The remaining part of the BH mass distribution follows directly
from the initial mass function and envelope removal through RLOF.

Additionally, we separately show these progenitors (and the corresponding
BHs) that will attain enough specific angular momentum  ($\tilde j \gtrsim
1$)
to produce a GRB. As mentioned before, it is found that only secondaries
form
such high rotating cores. Both distributions for this subpopulation, follow
closely their counterparts for potential GRB progenitors. Since only a small
fraction of secondaries gain high speeds and high specific angular momenta,
the numbers are much smaller for potential progenitors.
The majority of Population III collapsars ($\sim 98\%$) are formed from the
lower
mass end of the population III stars ($M_{\rm zams} \sim 10-100 \msun$), and
could be similar to Population I and II collapsars.  Only a small subset of
collapsars ($\sim 2\%$) originate from very massive stars ($M_{\rm zams}
\sim 200-500 \msun$) since such massive stars are hard to be accommodated
in binaries small enough that the synchronization would provide enough
angular
momentum to produce a GRB.

\subsection{Parameter Study}

Results of alternative evolutionary models are presented in Table~\ref{tab02}. 
In the following we discuss the effect of various parameters on our results. 

If the tides are only allowed to work on the envelope of the extended star
(giants with clear core and envelope structure) and not on its core (Mod05), 
the results remain unchanged. Since tides are the only effectively spinning
naked He
stars in close (post-CE) binaries, the tidal interactions prior to that stage
do  not play an important role in producing GRBs.  
Had we not allowed tidal interactions to be effective for naked He stars,
then the situation changes and of course we then do not produce any GRBs.
It is seen in a general case in which we do not allow any tidal
interactions (Mod02); no GRBs are formed, but almost the same number of potential
progenitors is found. Simply the stars are not spinning fast enough (no
tidal spin up) to satisfy any of specific angular momentum criteria we have
imposed.

We have assumed in our modeling that there is no angular momentum accretion
in RLOF events. That it is most probably not the case as the accreted mass
carries its own non-zero angular momentum. If angular momentum is
transferred
from one component to the other with mass, it may alter the spin of the
components.  
We have calculated a model (Mod14) in which we have allowed accreting stars to
maximally increase their rotational velocities (to Keplerian velocity). This
happens during stable RLOF (prolonged phase, mass accretion) when the
evolved primary transfers mass to the main sequence secondary. Main sequence
stars are assumed to rotate uniformly, and the entire star is spun up.
However, such a spin up has no effect on the predicted GRB formation
efficiencies. This is due to a simple fact that all GRB progenitors are
subject to tidal synchronization that sets the star speed in later
evolutionary stages. Therefore, no matter how much the star was spun up in
early evolutionary stages, once it is subject to tidal interactions it
will be slowed down to the speed set by the size of its orbit.  
We do not follow the corresponding spin down of the donors. Had it was
included it would not have altered the GRB formation efficiencies,
since the GRB progenitors originate from fast rotating helium cores that 
were spun up by synchronization in close binaries after mass transfer 
episodes were over. 

If we impose natal kicks for all compact objects (Mod03), some binaries are
disrupted and the GRB formation efficiencies ($\eta_{\rm i}$) decrease to
60-80\% of the standard model values. Although we have used quite large
kicks drawn from Maxwellian distribution with $\sigma = 265\ {\rm km\
s}^{-1}$
(Hobbs et al 2006; the most recent estimate for Population I stars), there
is only a rather small decrease in formation efficiency of GRBs. Once we
impose kicks many binaries are disrupted in SNe explosions. However, the
GRB progenitors evolve through one or two RLOF episodes and are found on
close orbits and therefore are hard to disrupt.    

The change of the minimum mass of a Population III star from 10 to $30
\msun$ (Mod04) increases the GRB formation efficiencies by factor of $\sim
3$. This is due to the fact that we assume that stars with $M_{\rm
zams}<19 \msun$ form NSs instead of BHs (unless these stars accrete
mass) and increasing of initial mass over the BH formation threshold
enhances the chances of forming a GRB progenitor.  The average stellar
mass in the simulation increases to $M_{\rm av,bin} \sim 88 \msun$,
and it is $\sim 2.5$ larger than in the standard model, therefore the
GRB detection rate remains basically unaffected as the two factors
cancel out (see eq.~\ref{Rone}). Since the binary fraction is the same
as in the standard model, the mass fraction $\eta_{\rm mass}$ does not
change either ($M_{\rm av,sin} \sim 105 \msun$ for this model).

We performed one calculation (Mod13) performed using our ``moderate rotator'' 
prescription for rotation speeds (see eq.~\ref{velzams}). 
The stars are initialized with speeds that are characteristic
of Population I stars ($\sim 200\ {\rm km\ s}^{-1}$). GRB formation
efficiencies does not change substantially, since the final spin up
for He cores making GRB progenitors does not depend on the star
initial spin, but rather on the size of the orbit of binary undergoing
synchronization.

We also perform a calculation with the alternative IMF for Population III
stars (Mod12); both component masses are drawn independently from a power-law
distribution with $\beta=-2.35$ within the same mass range as in the
standard
model ($10-500 \msun$). The resulting distribution of mass ratio, defined
now as mass ratio of the less massive component to the mass of the more massive
component), is flatter than in the standard model (see Figure~\ref{imf}).
The average mass of a star in binary and a single star is now the same and is
$M_{\rm av,bin}=M_{\rm av,sin}=28.9 \msun$, and we obtain $\eta_{\rm
mass}=0.18$. GRB formation efficiencies drop by factor of $\sim 2$ in this
model, however due to the lower average star mass in this simulation, the
{\em SWIFT} detection rates are not greatly affected (change by factor of
$\sim 0.75$).

Maximum increase of binary fraction (to $f_{\rm bin}=1$) may increase the
{\em SWIFT} detection rate by factor of $\sim 7$ (see eq.~\ref{etamass}).
However, such a high binary fraction in Population III seems very unlikely
due to the fact that primordial gas was much harder to fragment than
metal-rich gas of Population I and in particular binaries may have not
formed at all (e.g., Ripamonti \& Abel 2004).

We have also investigated the influence of adopted CE efficiency on our
results. With the decreased CE efficiency ($\alpha_{\rm ce} \times
\lambda=0.3$; Mod10) the GRB formation efficiencies change by factors $\sim 0.7$,
while with the increased CE efficiency ($\alpha_{\rm ce} \times \lambda=3$; Mod11) 
they change by factors $\sim 1.3$. Change of CE efficiency has an important
role on the post-CE orbital separation of a binary system. That in turn
plays
a crucial role for systems that are GRB progenitors (components need to be
close enough for tides to spin up naked He core of the secondary). If
post-CE orbit is too wide, He core is not spun up enough to form a GRB, and
if post-CE orbit is too tight, the binary components merge. The CE acts as
a simple orbital separation filtering, and changing the CE efficiency allows
CE to send different pre-CE binaries to the given post-CE separation range.
Therefore, as long as there is a wide range of pre-CE separations, change of
the CE efficiency will only select different binaries to fall within
specific
required post-CE orbit range (required, e.g., for GRB formation).      

Coupling of the He core with H envelope during giant phase is not important
for the formation of GRBs. Models with strong coupling ($X_{\rm cou1}=0.01$;
Mod06) as well as with fully decoupled He cores ($X_{\rm cou1}=1$; Mod07) 
generate the same numbers of
GRBs. Since the GRB progenitors are synchronous at giant stage (close
binaries) their cores rotate at the same speed as envelopes, and therefore
cores does not loose any angular momentum in the process.

In case of a very strong coupling of iron core to the surrounding He shell
($X_{\rm cou2}=0.01$; Mod09) the collapsing cores has 10 times less specific
angular
momentum then in the standard model, and the GRB formation efficiencies
change:
$\eta_2=0.0053$, $\eta_3=0.0006$, $\eta_4=0.0$ ($\eta_1$ is unchanged since
the similar potential progenitors are formed). Still the same population of
synchronized secondary helium stars make GRB progenitors. We note the
significant increase (factor $\sim 25$) in $\eta_2$, although the efficiency
is still relatively small.

\begin{figure}
\includegraphics[width=0.9\columnwidth,angle=0]{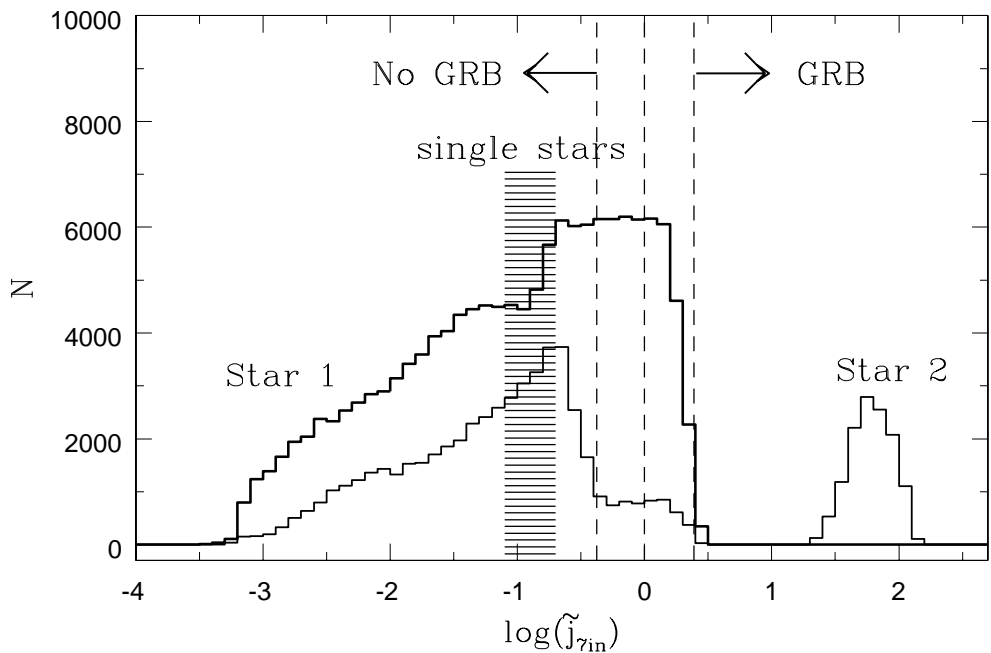}
\caption{
Specific angular momentum of potential GRB progenitors for model with
no angular momentum loss from the inner core (no coupling for iron core;
$X_{\rm cou2}=1$). Note that, in contrast to the standard model, both
primary and secondary can make a GRB.
Notation as in Figure~\ref{angm}.
}
\label{angm8}
\end{figure}

If we assume that an iron core is not coupled to the surrounding He
shell ($X_{\rm cou2}=1$; Mod08) then the core doesn't lose angular momentum
and is spinning fast. The specific angular momentum in this model is
presented in Figure~\ref{angm8}. Since specific angular momentum is
proportional to iron core speed, and the speed is higher by factor of
10 than in standard model, now both primaries and secondaries can
produce a GRB.  The subpopulation of secondary helium cores that are
spun up through tidal interactions and that are the sole GRB
progenitors in other models may in this evolutionary scenario have
specific angular momenta that are too high to produce a GRB ($20 <
\tilde j < 150$). The GRB formation efficiencies are following:
$\eta_1=0.1041$, $\eta_2=0.0242$, $\eta_3=0.0111$, $\eta_4=0.0067$. As
compared to standard model, $\eta_1$ does not change since it includes
all potential GRB progenitors (no constraint on specific angular
momentum), also $\eta_4$ is not affected too much as it includes only
very high angular momentum systems (synchronized helium stars as in
standard model), and the additional stars don't make the cut (see
Figure~\ref{angm8}). The highest increase (two orders of magnitude) is
observed for $\eta_2$ as it encompasses the largest fraction of
additional fast spinning cores (mostly originating from primaries),
and the moderate increase is found for $\eta_3$ which is doubled. Note
that the subpopulation of stars making GRBs in the standard model,
does not contribute to $\eta_2$ nor $\eta_3$ due to their very high
specific angular momentum. These are the highest GRB formation
efficiencies that we obtain within our modeling.

Although this assumption (no angular momentum loss from the inner core) may
be contrary to the expectations for single stars (Spruit 2002; Woosley \&
Heger 2006) it may potentially work for some stars in close binaries. For
binary components that are subject to strong tidal interactions, that
may make the entire star rotate synchronously, the iron core included. If
this is the case, the stars that go through synchronization are not subject
to angular momentum transport (uniform rotation) and then results of
this model should be considered as representative of binary population.
With the lack of observational evidence on the effectiveness of tides and
also
lack of theoretical work for Population III stars, we treat this as a major
uncertainty of our binary evolutionary model.

\subsection{Population III GRB Detectability}

Population III stars evolve very quickly so their redshift distribution
follows the star formation rate. For simplicity we assume that Population
III stars were formed between redshifts 10 and 30 and that the star
formation rate is in this epoch was $S=10^{-2} M_\odot {\rm Mpc}^{-3}$.  
For the calculation of the rate this is an optimistic assumption since the
actual redshift range for Population III stars is probably smaller.
For this choice of redshifts, we obtain $F(30)-F(10)=3.93$. The GRB
formation efficiencies are listed in Table~\ref{tab02} and we can use
equation~\ref{rate} to calculate SWIFT detection rate. We obtain $\sim
0.001-0.02 {\rm yr}^{-1}$ for standard model calculation, range resulting
from the different constrains on specific angular momentum used.  
Even for the model with  no angular momentum loss for iron cores, for which
we find the highest GRB formation efficiency ($\eta_2=0.0242$) {\em SWIFT}
detection rate goes only up to $\sim 0.1 {\rm yr}^{-1}$.  

This result has been obtained with rather conservative assumptions.
We have chosen a large binary fraction ($f_bin=0.1$) for Population III 
and a rather wide beaming fraction corresponding to the
typical beaming half angle of 11 degrees. We have also chosen 
a rather large value of the redshift range in which Population III
stars are formed. The star formation rate we have used can 
not be increased by much as well. Thus we consider our estimate 
as a very safe upper limit estimate for 
collapsars from the Population III binaries observable by SWIFT.

\section{Discussion}

We have modeled the evolution of Population III binaries using the
most up to date numerical models of such stars.  By focusing on binary
progenitors that both eject the hydrogen envelope and spin up the core
prior to collapse, we estimate the fraction of these systems that
produce potential GRB progenitors.  We find that the typical scenario
to form such a star produces GRBs from the star that is, at least
initially, less massive. It was found that significant fraction of
stars $\sim 0.1$ lose their H-rich envelopes through binary
interactions and are massive enough to form BH.  However, only a small
fraction of stars $\sim 0.01$ have been additionally spun up by tidal
interactions to a sufficient specific angular momentum to make a
GRB. Majority ($\sim 98\%$) of the collapsing cores that make GRBs
have relatively low masses and are in the range $\sim 5-50 \msun$, and
only a very few Population III collapsars are produce from massive
cores $\sim 150-300 \msun$.

We calculate the expected detection rate of Population III GRBs by SWIFT
and find that it is very low, $\sim 0.001-0.02 {\rm yr}^{-1}$ for our
standard evolutionary model. Even for the most optimistic model, in which we
do not allow angular momentum loss from the collapsing core, the rate is
rather low $\sim 0.1 {\rm yr}^{-1}$.  There are two main reasons for this.
First, the overall potential detection rate from Population III binaries is
relatively small. Even if we assume that all stars in Population III
binaries make a GRB, we would expect only $\sim 3 {\rm yr}^{-1}$, and this
only under optimistic assumptions on binary fraction and GRB collimation.
Second, it is difficult to spin the stars up through tidal interactions,
sufficiently enough to make a GRB. Only a small fraction of stars in
binaries ($\sim 0.01$) are subject to such a strong spin up.  
Bromm \& Loeb (2006) assumed that all the exposed He cores are spun up
sufficiently to make a GRB. This assumption led to an overestimate of high
redshift GRBs. In contrast, we find that there should be no detectable high
redshift GRBs from Population III binaries. This conclusion only weakly
depends on a number of evolutionary unknowns.
If, however, high redshift bursts are detected and are
identified with Population III stars, then our results argue against
this particular collapsar progenitor scenario for such bursts. One
should note that the expected rate of gravitational wave bursts from
coalescences of double black hole binaries originating in Population
III stars in the advanced LIGO is a much more promising way to
investigate the properties of Population III binaries (Belczynski et
al. 2004; Kulczycki et al.  2006).

The collapsar scenario requires that the star is stripped of its outer
layers in order for the jet to be able to break out. In the above
models we only consider stars stripped of hydrogen envelopes, however
some scenarios, and some observations, also require the removal of most 
of the helium envelope. At this stage the models are not advanced enough to
investigate this scenario in detail. However, we have estimated the
number of helium stars that fill the Roche lobes and hence may
potentially be stripped of their He-rich envelopes. Only a small
fraction of stars ($\lesssim 0.05$) in Population III binaries may
lose their He-rich envelopes and even if all of them are spun up
enough to produce GRBs the resulting rate is then only $\sim 0.15 {\rm
yr}^{-1}$.  One notable difference is that in this case most of the
bursts would come from the initially more massive star (Star
1). Further analysis, once better models of Population III stars are
available, should allow to investigate this issue.  Also, a component
merger, a new single star, may be spinning much faster than the
original components, and provide enough angular momentum to make
GRB. Such mergers are encountered in binary evolution during RLOF
phases.  Although we do not follow the evolution of such merger products,
their formation rate is also rather small; only about $\sim 0.05$
stars in Population III binaries merge before forming compact
remnants, resulting again in a rather small {\rm SWIFT} detection rate
of $\sim 0.15 {\rm yr}^{-1}$.  Another possibility of forming GRB in a
binary is a common envelope merger of a giant and a main sequence star
(Ivanova et al. in preparation; Podsiadlowski 2006, GRB symposium in
London). The inspiralling main sequence star supplies pure hydrogen to
the He burning shell of a giant, the resulting flash removes the
envelope of a giant. Also the in falling hydrogen-rich material may
spin up the Helium core of a giant. Although details of the model has
not been worked out in case of zero metallicity Population III stars,
we have estimated number of potential common envelope
configurations. Such common envelope mergers are found more frequently
$\sim 0.017$ than the GRBs formed through tidal spin up (see
Table~\ref{tab02}) but are still not numerous enough to warrant {\em
SWIFT} detection ($\sim 0.05 {\rm yr}^{-1}$), even if we assume that
all such events lead to a GRB (overestimate, since only very specific
initial conditions of an inspiral and stellar structure of binary
components may lead to a GRB; Ivanova, private communication).

We do not find enough GRB progenitors in Population III binaries to
warrant detection with {\em SWIFT}. If binaries exist in Population
III at all, they may provide potential gravitational radiation
signature, through mergers of their end products: massive double black
holes (see e.g., Kulczycki et al.  2006).  If GRBs are detected, an
ideal candidate might be the single star, high-mixing models proposed
by Yoon \& Langer (2005; see also Woosley \& Heger 2006).  Although 
this progenitor is unlikely to be the primary progenitor for GRBs at 
low redshift (Fryer 2006b), it could make a considerable contribution 
to the GRB population at high redshift. 
If all the stars were single in Population III, and if we have assumed 
that {\em all} of them produce a GRB, we would obtain {\em SWIFT} 
detection rate of $\sim 18\ {\rm yr}^{-1}$ (see eq.~\ref{rate}, with 
$\eta_i=1$ and $\eta_{\rm mass}=1$). However, during about 2 years of 
{\em SWIFT} observations there was no detection of an extreme redshift 
($z>10$) GRB. This, in principle, could put an upper limit on the 
Population III single star GRB model.

We acknowledge the support of KBN grant 1P03D02228, and KB thanks
Kasia Samson (Warsaw University) for help on this manuscript. Also we would
like to thank T.Abel, W.Dziembowski and B.O'Shea for comments on
initial spins and parameters of Population III stars.  Support for CLF
and AH was carried out under the auspices of the National Nuclear
Security Administration of the U.S. Department of Energy at Los Alamos
National Laboratory under Contract No. DE-AC52-06NA25396 and funded,
in part, by a NASA Grant SWIF03-0047, by the National Science
Foundation under Grant PHY99-07949 and by the DOE Program for Scientific
Discovery through Advanced Computing (SciDAC; DE-FC02-01ER41176).

\clearpage

\begin{deluxetable}{lccccc}
\tablewidth{400pt}
\tablecaption{GRB Formation Channels and Efficiencies\tablenotemark{a}}
\tablehead{Channel & Evolutionary Sequence\tablenotemark{b} & $\eta_1$ &
$\eta_2$ & $\eta_3$ & $\eta_4$}
\startdata
grb01 & MT1 GRB1 CE2 GRB2   & 0.0733 &  0.0002 & 0.0058 & 0.0063 \\
grb02 & CE1 GRB1 MT2 GRB2   & 0.0121 &    0    &   0    &   0    \\
grb03 & MT1 GRB1            & 0.0110 &    0    &   0    &   0    \\  
grb04 & All others          & 0.0077 &    0    &   0    &   0    \\
\hline
Total & All channels        & 0.1041 &  0.0002 & 0.0058 & 0.0063 \\
\enddata
\label{tab01}
\tablenotetext{a}{Formation efficiencies in case of no angular momentum
constraints ($\eta_1$); MacFadyen \& Woosley 1999 ($\eta_2$); this work
($\eta_3$); Podsiadlowski et al. 2004 ($\eta_4$).}
\tablenotetext{b}{Notation explained in \S\,3.1.}
\end{deluxetable}
\clearpage

\begin{deluxetable}{lccccl}
\tablewidth{350pt}
\tablecaption{GRB Formation Efficiencies: Parameter Study\tablenotemark{a}}
\tablehead{Model & $\eta_1$ & $\eta_2$ & $\eta_3$ & $\eta_4$ & Description\tablenotemark{b}}
\startdata
Mod01 & 0.1041 &  0.0002 & 0.0058 & 0.0063 & standard                               \\
Mod02 & 0.1103 &    0    &   0    &   0    & no tides                               \\
Mod03 & 0.0837 &  0.0001 & 0.0034 & 0.0037 & kicks                                  \\  
Mod04 & 0.2602 &  0.0008 & 0.0147 & 0.0146 & $M_{\rm zams,1} \geq 30 \msun$         \\
Mod05 & 0.1041 &  0.0002 & 0.0058 & 0.0064 & tides on envelope only                 \\
Mod06 & 0.1042 &  0.0002 & 0.0058 & 0.0064 & $X_{\rm cou1}=0.01$                    \\
Mod07 & 0.1039 &  0.0002 & 0.0058 & 0.0063 & $X_{\rm cou1}=1$                       \\
Mod08 & 0.1041 &  0.0242 & 0.0111 & 0.0067 & $X_{\rm cou2}=1$                       \\
Mod09 & 0.1041 &  0.0053 & 0.0006 &   0    & $X_{\rm cou2}=0.01$                    \\
Mod10 & 0.0831 &  0.0002 & 0.0037 & 0.0037 & $\alpha_{\rm ce} \times \lambda=0.3$   \\
Mod11 & 0.1295 &  0.0001 & 0.0082 & 0.0104 & $\alpha_{\rm ce} \times \lambda=3$     \\
Mod12 & 0.0471 &  0.0001 & 0.0020 & 0.0022 & alternative IMF                        \\
Mod13 & 0.1030 &  0.0002 & 0.0056 & 0.0062 & Pop I rotation                         \\
Mod14 & 0.1051 &  0.0002 & 0.0062 & 0.0068 & RLOF spinup                            \\
\hline
Range & .04--.26 &  0--.02 & 0--.01 & 0--.01 & all models \\
\enddata
\label{tab02}
\tablenotetext{a}{Formation efficiencies in case of no angular momentum
constraints ($\eta_1$); MacFadyen \& Woosley 1999 ($\eta_2$); this work
($\eta_3$); Podsiadlowski et al. 2004 ($\eta_4$).}
\tablenotetext{b}{Models explained in \S\,3.4.}
\end{deluxetable}
\clearpage

\clearpage

\clearpage

\clearpage

\clearpage

\clearpage

\clearpage

\clearpage

\end{document}